\documentclass[lettersize,journal]{IEEEtran}
\usepackage{bm}
\usepackage{color}
\usepackage{tabularx}
\usepackage{pifont}
\usepackage{bbding}
\usepackage{booktabs}
\usepackage{subeqnarray}
\usepackage{multirow}
\usepackage{makecell}
\usepackage{amsmath,amsfonts}
\usepackage{algorithmic}
\usepackage{algorithm}
\usepackage{array}
\usepackage[caption=false,font=normalsize,labelfont=sf,textfont=sf]{subfig}
\usepackage{textcomp}
\usepackage{stfloats}
\usepackage{url}
\usepackage{verbatim}
\usepackage{graphicx}
\usepackage{cite}
\usepackage{lipsum} 

\usepackage[colorlinks,
            breaklinks,
            pagebackref,
            linkcolor=red,
            anchorcolor=blue,
            citecolor=blue
            ]{hyperref}

\usepackage{xcolor}

\newcommand{\eg}{\textit{e}.\textit{g}.}
\newcommand{\etal}{\textit{et}~\textit{al}.}
\newcommand{\B}[1]{{#1}} 

\begin{document}

\title{DDistill-SR: Reparameterized Dynamic Distillation Network for Lightweight Image Super-Resolution}

\author{Yan~Wang,
        Tongtong~Su,
        Yusen~Li, 
        Jiuwen~Cao,
        Gang~Wang,
        and~Xiaoguang~Liu 
\thanks{Y. Wang, T. Su, Y. Li, G. Wang, and  X. Liu are with Nankai-Baidu Joint Lab, School of Computer Science, Nankai University, Tianjin, 300350, China, and also with the Tianjin Key Laboratory of Network and Data Security Technology, Tianjin, 300071, China. }
\thanks{J. Cao is with Machine Learning and I-health International Cooperation Base of Zhejiang Province, and Artificial Intelligence Institute, Hangzhou Dianzi University, China, 310018.} 
\thanks{Yusen Li is the corresponding author.}}



\maketitle

\begin{abstract}
Recent research on deep convolutional neural networks (CNNs) has provided a significant performance boost on efficient super-resolution (SR) tasks by trading off the performance and applicability. However, most existing methods focus on subtracting feature processing consumption to reduce the parameters and calculations without refining the immediate features, which leads to inadequate information in the restoration. In this paper, we propose a lightweight network termed DDistill-SR, which significantly improves the SR quality by capturing and reusing more helpful information in a static-dynamic feature distillation manner. Specifically, we propose a plug-in reparameterized dynamic unit (RDU) to promote the performance and inference cost trade-off. During the training phase, the RDU learns to linearly combine multiple reparameterizable blocks by analyzing varied input statistics to enhance layer-level representation. In the inference phase, the RDU is equally converted to simple dynamic convolutions that explicitly capture robust dynamic and static feature maps. Then, the information distillation block is constructed by several RDUs to enforce hierarchical refinement and selective fusion of spatial context information. Furthermore, we propose a dynamic distillation fusion (DDF) module to enable dynamic signals aggregation and communication between hierarchical modules to further improve performance. Empirical results show that our DDistill-SR outperforms the baselines and achieves state-of-the-art results on most super-resolution domains with much fewer parameters and less computational overhead. We have released the code of DDistill-SR at \url{https://github.com/icandle/DDistill-SR}.
\end{abstract}


\IEEEpeerreviewmaketitle

\section{Introduction}
\IEEEPARstart{S}{uper-resolution} (SR) is a classic computer vision problem, which aims to restore a high-resolution image from its degraded low-resolution counterpart~\cite{SISR}. As an essential task, SR has been widely used in various real-life applications, such as medical imaging~\cite{ZhuYL19}, surveillance imaging~\cite{zou2011very}, and real-time gaming media~\cite{chen2020learned}. 
Meanwhile, the demand for efficiency and robustness in these complex real-time applications continues to push forward the advancement of SR techniques.

\begin{figure}[!t]
  \centering
  \includegraphics[width=3.4in]{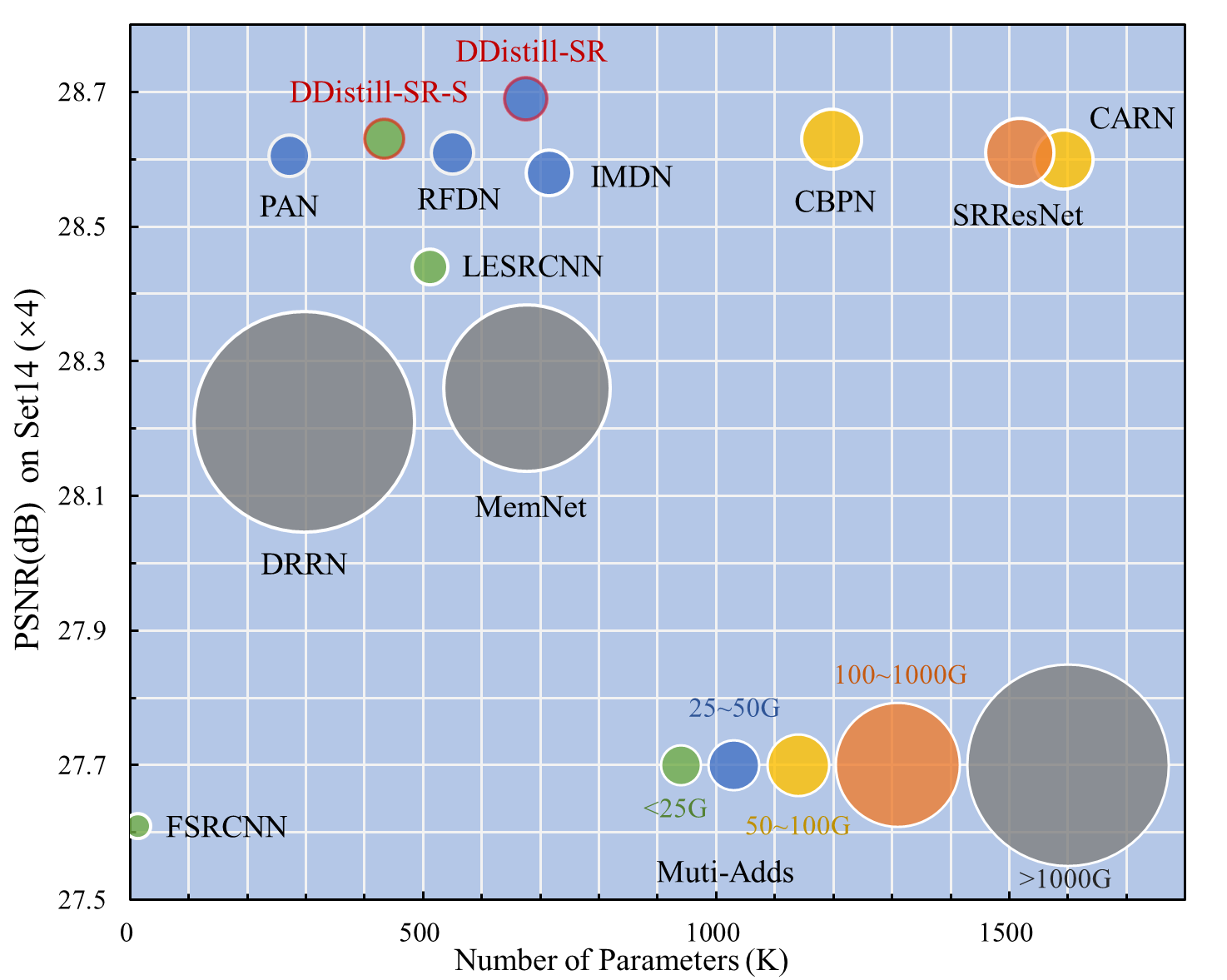} 
    \caption{Performance scores of the proposed method compared with existing lightweight super-resolution models on the Set14~\cite{set14} dataset. We set the magnification factor $\times 4$, and multi-adds are calculated with a $1280 \times 720$ SR shape. Best viewed in color.}
  \label{psnr_set14}
 \end{figure}

Despite its importance, developing a high-quality SR system is a non-trivial task due to its various ill-posed properties~\cite{yang2019deep}, \eg, notorious labeling and evaluation issues where one low-resolution (LR) input \B{can} correspond to multiple high-resolution (HR) images as the targets. 
The research community has made significant efforts to address these problems, which can be generally summarized as three strands: interpolation-based methods~\cite{keys1981cubic,zhang2006edge}, reconstruction-based methods~\cite{sun2008image,tai2010super} and deep learning-based methods~\cite{he2011single,romano2016raisr}. 
The first two methods are often fast and easy to develop for SR tasks, while their performance under challenging scenarios, \eg, using large magnification factors, is often inferior. 
In recent years, driven by the rapid development of \B{deep-neural-network-based} function approximators, CNNs have been the mainstream model to seek more effective SR solutions.
By learning the relationship between LR and HR image pairs, numerous CNN-based models~\cite{FSRCNN,CFSRCNN,RCAN} are proposed to restore satisfactory images with higher PSNR and SSIM~\cite{SSIM}.
\IEEEpubidadjcol
For instance, SRCNN~\cite{SRCNN} is a pioneer work to demonstrate the inherent advantage of the convolution layer in low-level vision tasks, which constructs a high-quality SR model surpassing most conventional algorithms.
The following works adopt various techniques, such as increasing the network's depth~\cite{VDSR,EDSR}, attention mechanism~\cite{ESA}, and advanced network structures~\cite{RDN}.
These methods often obtain a larger receptive field by utilizing a deeper or more complicated feature extraction procedure to summarize the inherent nonlinear mapping between low-and high resolution.
Nevertheless, simply enhancing feature extractors increases the cost of computation and memory in inference, \B{thereby} restricting the applications in scenarios with limited resources, such as embedded systems and mobile devices.

To develop an efficient and effective SR model, many lightweight networks with proper model \B{sizes} have been designed.
Based on the architectures of existing models \cite{wang2020deep}, these lightweight SR methods can be classified into the following three categories.
First, the recursive methods~\cite{DRCN} focus on leveraging the parameter-sharing blocks to reduce the model size and enlarge the receptive field of networks. 
Second, the residual/multi-path learning methods~\cite{SRResNet,s-LWSR} adopt multi-path or connections (\eg, residual or dense connection) among different blocks, to increase \B{the} feature flow connections and strengthen \B{the} model capability. 
Third, the layer-level modification methods leverage advanced layer designs to provide better structural awareness for SR, such as dynamic convolution operator~\cite{UDVD}, reparameterization operator~\cite{ECBSR}, and attention mechanism~\cite{ESA}.
However, the recursive learning methods with reusable modules only reduce parameters while \B{increasing the} computational complexity, which is unaffordable for real-world applications. The latter two strategies \B{attempt} to capture more informative features with powerful convolutional blocks or effective topology. However, these approaches tend to work independently and may be difficult to complement each other. \B{In addition}, another issue is that their fixed kernels and predefined trunk are inflexible to handle the input with diverse statistics, which may result in a monotonous feature expression and structural details loss. Therefore, this area \B{of research remains} relatively unexplored, which is essentially promising for performance enhancement.

In this article, we propose a lightweight super-resolution network called DDistill-SR \B{ which comprises} reparameterized dynamic units (RDU) and dynamic distillation fusion (DDF) modules \B{to better utilize the} immediate features under practical resource-restricted environments. Within RDU, we endow the convolution with dynamic adaptivity towards varied inputs and robust static kernels by integrating dynamic convolution and reparameterization strategies. Compared with other reparameterization and dynamic convolution, our method with convolution’s inference \B{equally} captures more representative static and dynamic features by dynamically fusing several parallel reparameterizable blocks. For DDF, we develop parallel dynamic feature distillation and fusion pipelines to refine the static and dynamic features jointly. Unlike \B{the} previous information distillation design~\cite{IMDN} that focuses on pruning static features, the proposed DDF reuses the intermediate dynamic features yielded by RDUs at the sub- and full-network level.
We compare DDistill-SR with several state-of-the-art methods, such as LESRCNN~\cite{LESRCNN}, PAN~\cite{PAN}, and RFDN~\cite{RFDN}. The results in Fig.~\ref{psnr_set14} show that our networks achieve a better trade-off \B{among} the restoration accuracy, parameters, and multi-add operations.     

Overall, the main contributions of our paper are three-fold: 
\begin{itemize}
  \item [1)] 
   We propose a novel reparameterized dynamic unit (RDU) to extract the robust static and adaptive dynamic features by dynamically combining parallel reparameterizable blocks for better layer-wise representation. The RDU is trained with complex structures and deployed with convolution inference, \B{therefore} ensuring efficiency. Experiments show that the RDU module significantly improves the performance of multiple super-resolution networks with a negligible increase in parameters and calculations.
  \item [2)]
  We propose a dynamic distillation fusion (DDF) module to maximize the \B{effects} of the hierarchical dynamic information by refining dynamic features step-by-step for network-level enhancement. Within the DDF, the informative dynamical features \B{can be selectively retained and distilled throughout the entire feature extraction process}, which \B{can} improve both restoration fidelity and accuracy.
  \item [3)]
  We present a reparameterized dynamic distillation super-resolution network (DDistill-SR) to improve the immediate feature representation by applying RDU and DDF for efficient SISR. The qualitative and quantitative results demonstrate that our method aptly balances the restoration performance and computational complexity.
  
\end{itemize}

The \B{remainder} of this paper is organized as follows. In Section~\ref{section:related_work}, we review the related works on image super-resolution and enhancement techniques on CNN. The design of DDistill-SR is presented in Section~\ref{section:method}. In Section~\ref{section:experiment}, we evaluate DDistill-SR and compare it with various state-of-the-art approaches. Finally, we conclude this work in Section~\ref{section:conclusion}.

\section{Related Work}\label{section:related_work}

\subsection{Single-Image Super-Resolution} \label{subsecion:SISR}
As the first CNN-based SR method, SRCNN~\cite{SRCNN} uses a 3-layers end-to-end architecture to fit nonlinear mapping, which outperforms most traditional SR methods.
Following SRCNN, Dong~\etal~\cite{FSRCNN} has proposed FSRCNN, which uses the \B{post-deconvolution} up-sampling layer and $ 1 \times 1$ convolution to diminish computations and parameters. VDSR~\cite{VDSR} deepened the network with 20 convolution layers to obtain a vast receptive field for adapting various upscaling factors, which has \B{massively improved the} performance. However, increasing the depth of these models aggravates the convergence difficulty due to their structural defects, which \B{restricts} both inference efficiency and model performance.

To address \B{this} issue, residual-learning methods, \eg, EDSR~\cite{EDSR}, have stacked more convolution layers to learn global and local residuals. Since then, several SR models with larger receptive fields have been proposed to improve \B{the reconstruction quality}. However, a deeper network means more parameters and \B{heavier} load. From this perspective, recursive learning has been applied to accelerate convergence and reduce model size based on shared weights and skip connections. For example, DRCN~\cite{DRCN} employs a single convolution layer to increase the depth of the recursion layer, and Lai~\etal~\cite{LapSRN} combined \B{the} cascade Laplacian pyramid architecture and Charbonnier loss to obtain a stable training process. However, the recursive learning methods have much higher computational complexity because they have to repeatedly use the same block to ensure a good feature representation. Another effective strategy for enhancing the restoration performance is to maximize the immediate feature extraction efficiency by employing the powerful attention mechanism or convolution layer. For example, RCAN\cite{RCAN} used channel attention to obtain more informative features by exploiting the channel-wise interdependencies, and HAN\cite{HAN} adopted holistic attention to learn the channel and spatial correlation of each layer. However, these networks cannot be applied to lightweight scenarios since they have hundreds of layers.

In order to achieve a better trade-off between the model complexity and performance, many competitions~\cite{AIM2019,AIM2020} and works~\cite{MAFFSRN,LESRCNN} have been carried out for lightweight SR. CARN-M and CARN~\cite{CARN} use local and global cascading architectures, \B{and they achieve} more efficient image recovery and video recovery on mobile devices. Hui~\etal~\cite{IDN} integrated enhancement and compression units into the information distillation network to achieve efficient inference time. Tian~\etal~\cite{CFSRCNN} proposed CFSRCNN with a cascaded structure and an efficient feature extractor to prevent potential training instability and performance drop. However, the most critical information in \B{these} intermediate features is not highlighted in the methods, which would cause inefficient feature extraction and inferior performance.

The latest works absorbed multiple strategies for better representation capability to address the above issue. For example, Zheng~\etal~\cite{IMDN} adopted the information multi-distillation block~(IMDB) and leveraged contrast-aware channel attention in the final stage of the block to emphasize channel attributes of features generated by progressive refinement. PAN~\cite{PAN} integrated pixel attention and self-calibrated convolution to gain competitive super-resolution results with only 272k parameters. RFDN~\cite{RFDN} further improved IMDB by using a lighter feature distillation function and a more powerful enhanced spatial attention (ESA)~\cite{ESA} to achieve higher efficiency\B{, and it} won first place in the AIM 2020 efficient SR challenge. In addition, Tian~\etal~\cite{ACNet_SR} introduced \B{an} asymmetric CNN (ACN) to enhance the square convolution kernels in \B{the} horizontal and vertical directions and develop the multi-level feature fusion mechanism to achieve an excellent trade-off between performance and complexity.

\begin{figure*}[!t]
  \centering
  \includegraphics[width=7in]{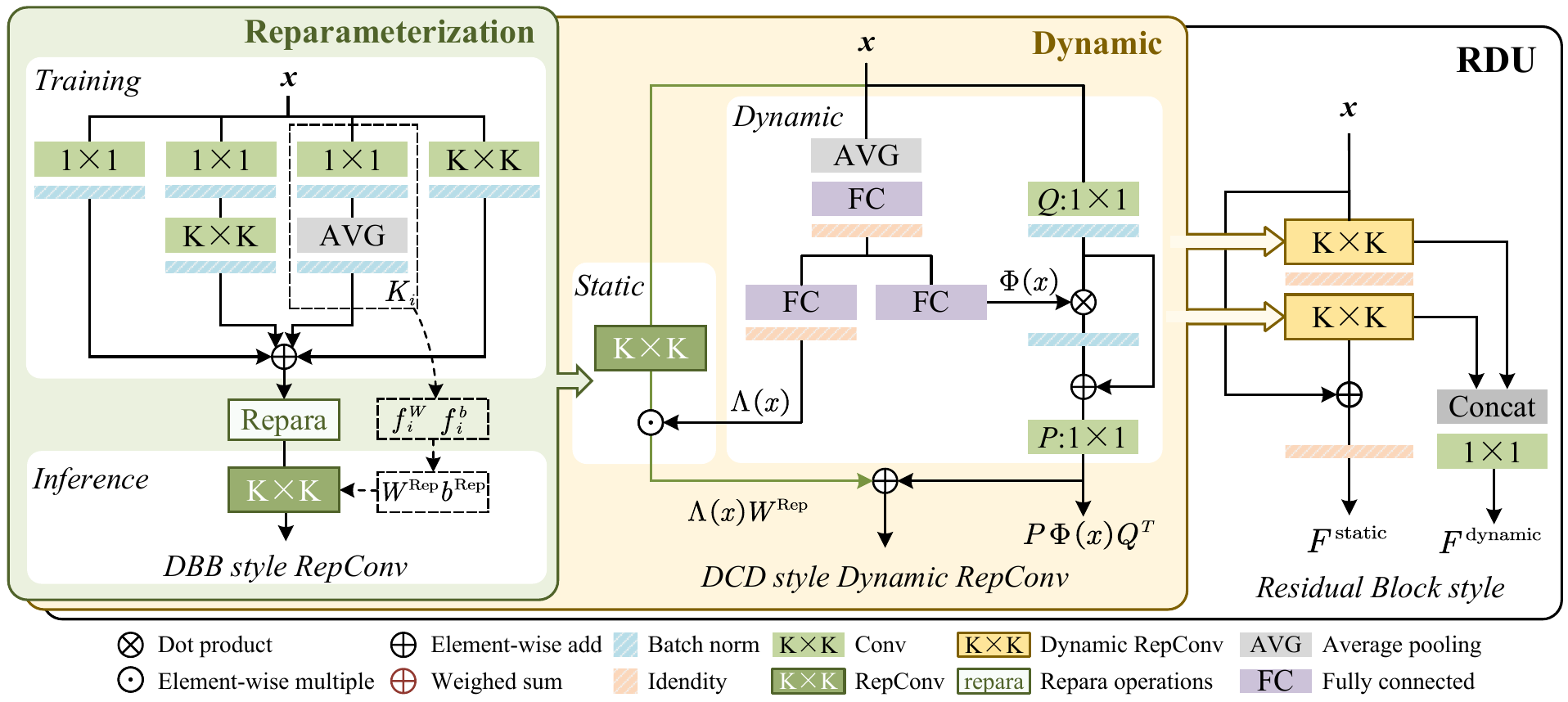} 
  \caption{Illustration of the implementation of RDU with residual block structure. RDU can be decomposed into two parts: reparameterization and dynamic. For the reparam strategy, the light green block shows the structure of DBB-style RepConv, which is trained in multi-branches and deployed as a static convolution. For dynamic, the light yellow block reveals the architectures of DCD-style Dynamic RepConv, which strengthen the static branch with RepConv.} 
  \label{fig_sim}
 \end{figure*}
 
\subsection{Convolutional Block Design}\label{subsecion:efficient}
Recently, lightweight convolution blocks have been widely used in SR, which are more powerful and flexible for deployment in real life~\cite{Dilated_Conv,Deformable_Conv,Xception,AC}. We summarize two representative lightweight convolution block methods, model parameterization, and dynamic convolution, which are used in our work. 

Model reparameterization converts a trained heavy block into a simple one, which has been widely adopted in CNN-based models. For example, \B{the} asymmetric convolution block (ACB)~\cite{AC} combines an asymmetric convolution skeleton and \B{a} convolution baseplate into a standard convolution, which greatly ameliorates the performance of CNNs. \B{The} RepVGG~\cite{RepVGG} stacks many reparameterized convolutions, which achieves higher accuracy and faster speed on classification tasks.
Driven by the reparameterization technique, the lightweight SR performance has been sufficiently promoted. FIMDN~\cite{AIM2020} leverages ACB to enhance the IMDN and shows the potential of reparameterization to improve the SR performance without changing the model inference.
Ding~\etal~\cite{DBB} devises an inception-like Diverse Branch Block (DBB) to extract multiple-path features in the training stage, which gains remarkable improvement and broader application on numerous architectures. 
Inspired by RepVGG and DBB, Zhang~\etal~\cite{ECBSR} has introduced a novel reparameterized block, \B{called the} edge-oriented convolution block (ECB), to balance hardware efficiency and restored visual quality in real-time mobile applications.
Nevertheless, these reparameterized methods greatly increase the training complexity, which limits their application in deeper networks.

\B{Unlike} reparameterization, dynamic convolution improves \B{the} performance over standard convolution by using additional parameters and calculations. For example, conditionally parameterized convolutions (CondConv)~\cite{CondConv} aggregates multiple convolution kernels \B{and} delivers outperformance in classification missions. Similarly, DYConv~\cite{DYConv} aggregates kernels according to dynamic attention to achieve higher efficiency. Then, Chen~\etal~\cite{DRConv} \B{introduced} spatial awareness into dynamic convolution by learning guided masks to further improve performance. Gaussian dynamic convolution (GDC)~\cite{GDC} \B{was} introduced to collect contextual information by randomly sampling the spatial area according to the Gaussian distribution offsets.
However, these methods contain a large number of redundant parameters, which hinders their deployment on mobile devices. 
To \B{improve} the accuracy \B{while maintaining a} proper model size, Li~\etal~\cite{DCD} leverages dynamic convolution decomposition (DCD) to replace onerous dynamic attention over channel groups, which achieves significant growth in accuracy and convergence rate.
Recently, dynamic convolution has been widely applied in dynamical magnification and degradation super-resolution tasks. 
UDVD~\cite{UDVD} has proposed 2 types of dynamic convolution to restore both synthetic and real images under variational degradation. 
Wang~\etal~\cite{ArbSR} has designed a plug-in adaptive upsampling module with a scale-aware convolution layer to achieve a scale-arbitrary SR scheme.

Although \B{these} convolution blocks have achieved breakthroughs in the SR field, there still exists a huge promoting space for obtaining better generality and effectiveness. In this paper, we combine the advantages of \B{the} dynamical convolution and reparameterization to propose a more efficient block, which improves \B{the} information capture capability. 

\section{Methodology}\label{section:method}
In this section, we present the technical details of the proposed DDistill-SR network. 
Overall, it consists of two main components: a reparameterized dynamic unit (RDU) which serves as a plug-in replacement for the conventional blocks; and a reparameterized dynamic feature distillation block (RepDFDB) which employs multiple-step distillation and local fusion module to generate informative feature maps. 
Then, we introduce \B{the formulation of} these two components and demonstrate how to stack them to build our DDistill-SR.

\subsection{Reparameterized Dynamic Unit (RDU)}\label{section:method-RDU}
We formulate a reparameterized dynamic unit (RDU) \B{by} integrating the structural reparameterization and dynamic convolution techniques, which forms a more effective plug-in replacement over the original convolution block.
Since these two strategies are convolution-based reformation, we first give the mathematical notions of a convolution layer to clearly explain RDU as \B{follows}:
\begin{equation}
  h (\bm{x}) = \bm{W}\ast\bm{x} + \bm{b},
  \label{conv}
\end{equation}
where $\bm{x}$ \B{is} the input features and $h (\bm{x})$ \B{is} the output features with $C$ and $D$ channels. $\ast$ denotes the convolution operation.
$\bm{W} \in \mathbb{R}^{D\times C\times k\times k}$ and $\bm{b} \in \mathbb{R}^{1\times D\times 1\times 1}$ represent the weight and bias for the $k\times k$ convolutional kernel, respectively.
To achieve the best feature extraction, we adopt a dynamic convolution to obtain \B{the} dynamic weight $\bm{W}(\bm{x})$. For $K$ static expert kernels, the dynamic kernel is their weighted sum determined by an input-relevant attention mechanism $\pi_k(\bm{x})$. This information fusion process can be encapsulated as:
\begin{equation}
  \bm{W}(\bm{x}) = \sum_{k = 1}^{K}\pi_k(\bm{x})\bm{W}_k,
  \label{dyconv} 
\end{equation}

\B{In addition to} enhancing \B{the} convolution \B{via} dynamical fusion, we leverage structural reparameterization to get preferable static $\bm{W}^{\rm {Rep}}$ and $\bm{b}^{\rm {Rep}}$.
It converts one complex architecture, \eg, an inception-style block, to a static convolution block \B{by} transforming its parameters. 
This process does not engage additional model parameters for inference through static fusion.
We denote the reparameterized convolution as RepConv, which includes RepVGG, DBB, and ECB. Given a $N$ branch reparameterizable block, the procedure of using $f^W$ and $f^b$ to convert its all trainable and fixed parameters $\bm{K}$ to a $k\times k$ static convolution can be summarized as follows:
\begin{subequations} 
  \begin{align}
   \bm{W}^{\rm {Rep}} &= \sum_{i=1}^{N}{f}^{W}_i(\bm{K}_i),\label{rep_W} \\
   \bm{b}^{\rm {Rep}} &= \sum_{i=1}^{N}{f}^{b}_i(\bm{K}_i),  \label{rep_b} 
\end{align}
\end{subequations}
where $\bm{W}^{\rm {Rep}}$ and $\bm{b}^{\rm {Rep}}$ are reparameterized matrices of the kernel and bias for \B{the} static $k\times k$ convolution. In Fig.~\ref{fig_sim}, we present an example of DBB-style RepConv, which can be replaced by any other reparameterizable block, \eg, RepVGG.

On the premise that both dynamic and static enhancements are accessible, the problem comes:``How to best integrate these two techniques?". 
Although directly replacing static kernels with reparameterized kernels is available and easy to implement, the parameter number is apparently unacceptable for a lightweight network when the growth rate of parameters number is $K\cdot N$.
Following DCD~\cite{DCD}, we leverage matrix decomposition to simplify the dynamical convolution in Eq.~\ref{dyconv}.
Assuming the dynamic convolution operation maintains the channel number of $C$, the $k \times k$ size dynamic kernel in DCD can be calculated in the following manner:
\begin{equation}
  \bm{W}(\bm{x})=\Lambda(\bm{x}) \bm{W} + \bm{P} \Phi (\bm{x}) \bm{Q}^T, 
  \label{dcd} 
\end{equation}
where $\Lambda (\bm{x})$ is a $C \times C$ diagonal matrix related to \B{the} squeeze-and-excitation (SE) module~\cite{RCAN}. 
$\bm{Q}$ is a $ C \times L$ matrix \B{that reduces the number of} input channels from $C$ to $L$. Correspondingly, $\bm{P}$ is another matrix to broaden the width from $L$ to $C$. 
$\Phi (\bm{x})$ is a $L\times L$ tensor decided by individual input $\bm{x}$, and it dynamically performs feature fusion in latent space. 
Different from the classification task, the maximum of the latent space $L$ is artificially enlarged to $\frac{C}{2}$ to ensure ample dynamical information.
\B{In addition,} the dynamic weight $\bm{P} \Phi \bm{Q}^T$ can be separated to generate an individual output, which is informative in feature representation. 
According to this, we design a special module to collect and reuse them as an effective feature representation supplement.
Based on these, we combine structural reparameterization and dynamic convolution into a convolution process to construct an advanced Dynamic RepConv that maintains the running-time structures to make the best use of each convolution operation by:
\begin{equation}
    \hat{h}(\bm{x}) = \Lambda(\bm{x}) \bm{W}^{\rm {Rep}} \ast \bm{x} + \left(\bm{P} \Phi (\bm{x}) \bm{Q}^T\right) \ast \bm{x} + \bm{b}^{\rm {Rep}},\label{DRConv}
\end{equation}
where $\hat{h}(\cdot)$ represents the advanced Dynamic RepConv used in RDU.
$\bm{W}^{\rm {rep}}$ and $\bm{b}^{\rm {rep}}$ are the reparameterized weight and bias converted from complicated reparameterizable blocks. 
 
Finally, we construct plug-in RDUs with the modified Dynamic RepConvs and potential arrangements (\eg, residual block style).
Each RDU consists of two outputs: the static features extracted by the normal block and the additional dynamic features fused by \B{the} shallow fusion part. As shown in Fig~\ref{fig_sim}, the normal block is a general residual block but uses powerful Dynamic RepConv to extract static features $F^{static}$. The shallow fusion part leverages concatenation and convolution operations to integrate two $\bm{P} \Phi \bm{Q}^T$ into dynamic residual $F^{\rm{dynamic}}$.
The \B{entire} process of residual block style RDU can be given by:
\begin{equation}
\begin{aligned}
        F^{\rm{static}} &=  \hat{h}_1\left(\hat{h}_2\left(\bm{x}\right)\right)+\bm{x},\\    
        F^{\rm{dynamic}}& = h \left(C\left(\left(\bm{P} \Phi \bm{Q}^T\right)_1,\left(\bm{P} \Phi \bm{Q}^T\right)_2\right)\right), \label{RDU}  
\end{aligned}
\end{equation}
where $\hat{h}_1$ and $\hat{h}_2$ are two Dynamic RepConv in residual block. $\left(\bm{P} \Phi \bm{Q}^T\right)_i$ is the $i$-th dynamic residual generated by $\hat{h}_i(\cdot)$. $h(\cdot)$ and $C(\cdot)$ represent the $1\times 1$ convolution and the concatenation operation along the channel dimension respectively.

\subsection{Reparameterized Dynamic Feature Distillation Block} \label{Subsection:block}

\begin{figure}[!t]
  \centering
  \includegraphics[width= 3.3in]{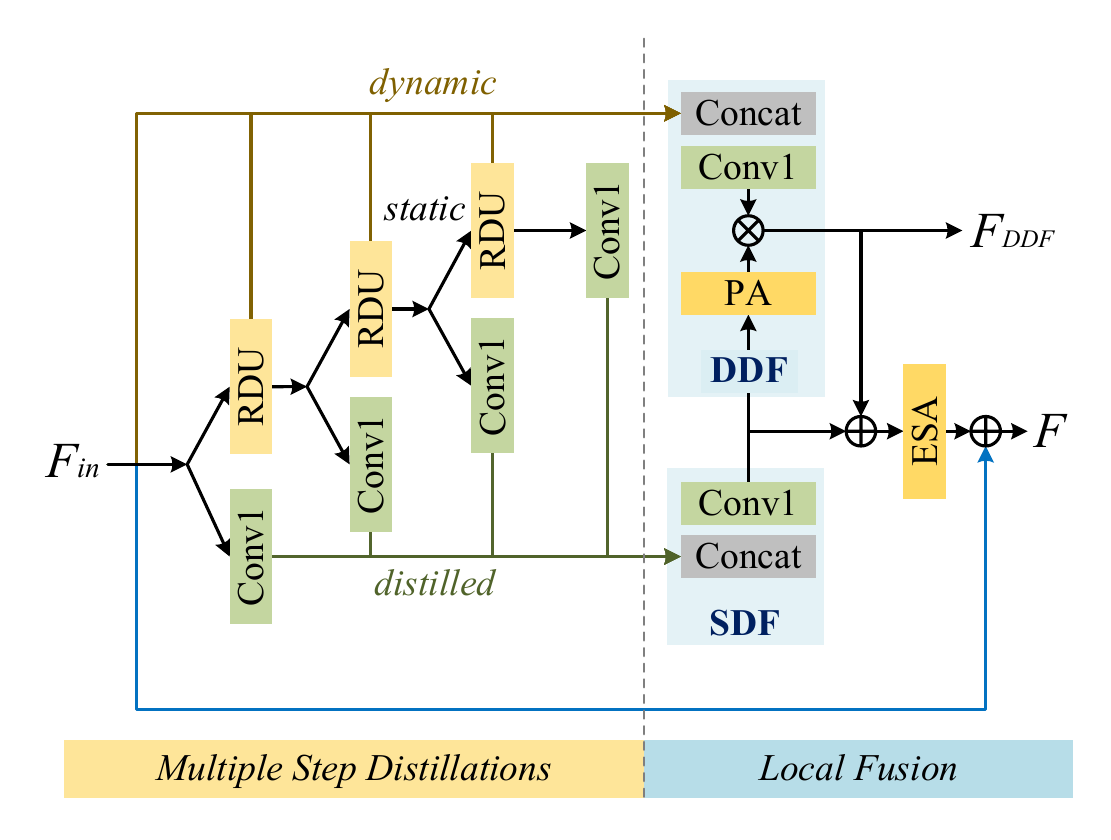} 
  \caption{Architecture of our proposed reparameterized dynamic feature distillation block (RepDFDB). Here, the ``RDU” denotes the proposed unit in~\ref{section:method-RDU}, and ``SDF” and ``DDF" represent static/dynamic distillation fusion along channel dimension. We use black/brown/green \B{lines} to show the flowing of static/dynamic/distilled features, respectively.}  
  \label{fig_RepDFDB}
 \end{figure}

\begin{figure*}[!t]
  \centering
  \includegraphics[width=7.1in]{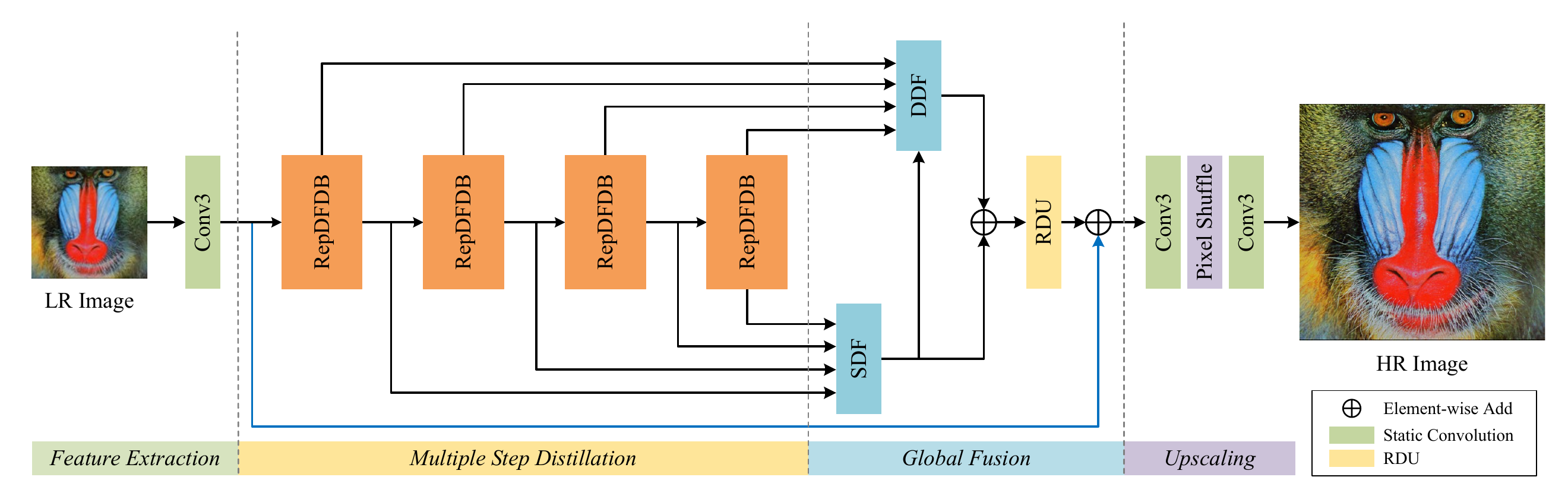} 
  \caption{Overview of the architecture of our proposed DDistill-SR network. DDistill-SR has 4 RepDFDB stages and conducts upscaling via sub-pixel after the global DDF enhancement. DDistill-SR-S shares a similar structure except for the DDF module.}  
  \label{fig_net}
 \end{figure*}
 
Inspired by information distillation SR networks~\cite{IMDN,RFDN}, we propose a novel feature distillation processing block, the reparameterized and dynamic feature distillation block (RepDFDB), to reject dynamic information into the distillation process. The design of RepDFDB is shown in Fig.~\ref{fig_RepDFDB}.
Particularly, RepDFDB has two parts: the multiple-step distillation using RDU and the local fusion module. 
As labeled in the left part of Fig.~\ref{fig_RepDFDB}, the multiple-step distillation module adopts the proposed RDU block and a convolution layer to extract static, dynamic, and distilled features for subsequent refinement steps.
For each step, RDU is employed to calculate $n$-th $F^{\rm{static}}_n$, $F^{\rm{dynamic}}_n$, and $1 \times 1$ convolution layer is devised for $F^{\rm{distilled}}_n$. 
Given the input feature $F_{\rm{in}}$, the multiple-step distillation procedure can be given by:
\begin{equation}
\begin{aligned}
  F^{\rm{static}}_1, F^{\rm{dynamic}}_1, F^{\rm{distilled}}_1 &= RDU_1(F_{\rm{in}}), h(F_{\rm{in}}),\\ 
  F^{\rm{static}}_2, F^{\rm{dynamic}}_2, F^{\rm{distilled}}_2 &= RDU_2(F^{\rm{static}}_1), h(F^{\rm{static}}_1),\\
  F^{\rm{static}}_3, F^{\rm{dynamic}}_3, F^{\rm{distilled}}_3 &= RDU_3(F^{\rm{static}}_2), h(F^{\rm{static}}_2),\\
                     F^{\rm{dynamic}}_4, F^{\rm{distilled}}_4 &= F_{\rm{in}}, h(F^{\rm{static}}_3),\\
\end{aligned}
  \label{multisplit} 
\end{equation}
where $F^{\rm{static}}_n$, $F^{\rm{dynamic}}_n$ and $F^{\rm{distilled}}_n$ are $n$-th static, dynamic and distilled features, respectively. The $RDU_n$ represents RDU according to Eq.~\ref{RDU},

Existing methods only leverage static information in the distillation fusion process. In contrast, our local fusion module in RepDFDB involves both static and dynamic distillation merging \B{processes} to obtain further improvements.
For conventional static distillation fusion (SDF), $F^{\rm{distilled}}_n$ are concatenated and then processed by a $1\times1$ convolution layer to perform channel fusion as follows: 
\begin{equation}
 F_{\rm{SDF}} = h (C(F^{\rm{distilled}}_1,F^{\rm{distilled}}_2,F^{\rm{distilled}}_3,F^{\rm{distilled}}_4)),
\label{merge1} 
\end{equation}
where $F_{\rm{SDF}}$ is the static fused feature map in the local level.
For our proposed dynamic distillation fusion (DDF), as depicted in Fig.~\ref{fig_RepDFDB}, we merge the $F^{\rm{dynamic}}_n$ of Eq.~\ref{RDU} to calculate block-wise dynamic residue, but also refine the fused features with pixel-attention (PA)~\cite{PAN} mechanism $f_{\rm{PA}}(\cdot)$.
The attention map of the static result is used to calibrate the dynamic fusing procedure.
In general, the \B{entire} process of DDF can be formulated as:
\begin{equation}
\begin{aligned}
    F_{\rm{DDF}}&=f_{\rm{PA}}(F_{\rm{SDF}})\cdot\\ h&(C(F^{\rm{dynamic}}_1,F^{\rm{dynamic}}_2,F^{\rm{dynamic}}_3,F^{\rm{dynamic}}_4)),  
\end{aligned}
\label{ddf2}
\end{equation} 
where $F_{\rm{DDF}}$ is the refined dynamic residual \B{at the local level} and also is the dynamic output for RepDFDB.
Finally, we apply the enhanced spatial attention (ESA)~\cite{ESA} and block-scale skip connection to strengthen \B{the} high-frequency spatial perception. Therefore, the final static output of RepDFDB can be computed by:
\begin{equation}
  F = f_{\rm{ESA}}(F_{\rm{SDF}}+F_{\rm{DDF}})+ F_{\rm{in}}, 
  \label{dfdb_out} 
\end{equation}
where $F_{\rm{in}}$ denotes the input features, and $f_{\rm{ESA}}(\cdot)$ is the enhanced spatial attention module to enhance spatial context.

\subsection{Network Architecture}
The architecture of DDistill-SR is shown in Figure~\ref{fig_net}.
Overall, there are four consecutive sub-modules to learn meaningful features from the original inputs: Feature Extraction module, Multiple Step Distillation module, Global Fusion module, and Upscaling module.

The LR images are first processed with the Feature Extraction module which utilizes a single $3\times3$ convolution layer to generate elementary features.
Given the input LR images $I_{\rm{LR}}$ with the shape of $N\times 3\times H\times W$, the low-level extracted feature $F_{\rm{FE}}$ has the same height $H$ and width $W$ but larger channel number $C$.

Then, features are fed into the Multiple Step Distillation and Global Fusion module for intermediate feature extraction and fusion. These modules can \B{be considered} as extensive versions of the corresponding components in RepDFDB because both designs utilize hierarchical feature detection and fusion modules according to Eq.~\ref{multisplit} and Eq.~\ref{merge1}-\ref{dfdb_out}, respectively. 
In \B{the} global Multiple Step Distillation, RepDFDB is devised to replace the RDU and $1\times1$ convolution, where $F$ calculated by Eq.~\ref{dfdb_out} is used as the static and distilled feature maps, and $F^{\rm{DDF}}$ in Eq.~\ref{merge1} as the dynamic feature map.
In the Global Fusion module, we use the RDU instead of ESA to enhance the high-frequency information with fewer parameters.
Hence, the intermediate features can communicate among the block and network by progressively refining the static and dynamic information.

Finally, the feature yielded by the Global Fusion module is sent to the Upscaling module, \B{which consists} of non-parametric sub-pixel operation and several convolutional layers for all scales in particular on light and efficient reconstruction. For \B{the} $\times s$ upscaling task, we first reduce the feature channels to $3s^2$ in \B{the} first convolution layer, \B{then we} convert the feature maps from \B{the} LR space to \B{an} HR image \B{using} a pixel shuffle layer. Moreover, we add the second convolution layer as a filter to improve the visual verisimilitude.

\section{Experiments}\label{section:experiment}

\subsection{Datasets and Metrics}
We conduct the whole training process on DIV2K~\cite{div2k} and Flick2K~\cite{EDSR}, which are widely used in multiple SR tasks. Specifically, 3450 high-quality RGB images are included in the training set.
In order to verify the generality and effectiveness of our DDistill-SR model, three common degradation methods are applied to obtain the LR-HR image pairs: \textbf{BI}, \textbf{BD}, and \textbf{DN} for imitating LR images in the real world.
Then, we investigate the performance of different SR algorithms on five well-known datasets: Set5~\cite{set5}, Set14~\cite{set14}, B100~\cite{B100}, Urban100~\cite{Urban100}, and Manga109~\cite{manga109}, each of which has different characteristics.
Referring to most SR methods, we utilize the peak signal-to-noise ratio (PSNR) and structural similarity (SSIM)~\cite{SSIM} as the quality evaluation metrics on the $Y$ channel of the $YCbCr$ space.

\subsection{Implementation Details}
In general, we extract LR-HR patch pairs with the LR size of $64\times 64$ from the preprocessed images in DIV2K and Flickr2K. Regarding LR-HR pair preparation, the commonly used down-sampling approach, bicubic interpolation (\textbf{BI}), is applied under magnification factors $\times2$, $\times3$, and $\times4$ for the basic tests. In addition, the blur-down-sampled (\textbf{BD}) method is utilized to blur HR images with a $7\times 7$ Gaussian kernel and standard deviation $\sigma = 1.6$, then downscaled with bicubic interpolation to generate $\times3$ datasets. Thirdly, a more complicated method is used to test SR restoration in extreme cases. The down-sampled-noisy (\textbf{DN}) method processes HR images with bicubic downsampling followed by additive 30\% Gaussian noise. During training, we rotate the images by 90°, 180°, 270° and flip them horizontally for data augmentation. The minibatch size is set up to 64 and \B{the} ADAM optimizer ($\beta_1 = 0.9$, $\beta_2 = 0.999$) \cite{ADAM} is applied to update model parameters. Moreover, the cosine annealing learning scheduler is utilized to achieve faster convergence. The learning rate is initialized as $5\times10^{-4}$, and the minimum learning rate is set as $1\times10^{-7}$. The cosine period is 250k iterations, and four periods are conducted in the process. For convenience, we use the $\times2$ model as the pre-trained model to train the $\times3$ and $\times4$ models. The final models are fine-tuned before and after the reparameterization operation. We implement the final DDistill-SR with 56 channels and RDU with the base design. In terms of loss functions, the $L_1$ loss is utilized to optimize the DDistill-SR model, and $L_2$ loss is used to fine-tune. All experiments are conducted on the Pytorch framework with a single RTX 3090 GPU.

\begin{table}[!t]
  \centering
  \setlength\tabcolsep{2.2pt}
  \small
  \caption{Investigations of \B{the} reparameterization performance on B100~\cite{B100} testset with $\times 4$ \textbf{BI} degradation.}
  \begin{tabular}{lrcccc}
    \toprule
    Model & Params &  MAdds  &Runtime& PSNR/SSIM \\
    \midrule 
    w/o Rep & 675K & 32.62G   & 0.0137s &  27.5946/0.737121\\ 
    w/ Rep (training)  & 1173K & 60.80G   & 0.0279s &27.6455/0.738465\\ 
    w/ Rep (inference) & 675K & 32.62G   & 0.0137s &  27.6457/0.738467\\ 
    \bottomrule
  \end{tabular}
\label{tab:ablation_repara}
\end{table}

\begin{table}[!t]
  \centering
  \caption{Investigations of RDU on \textbf{BI} benchmarks ($\times 4$). The Params and MAdds are calculated with training models. }
  \small
  \setlength\tabcolsep{3.8pt}
  \renewcommand{\arraystretch}{1.1}
  \begin{tabular}{cccrcc}
    \toprule
    \multirow{2}*{Model} &  \multicolumn{2}{c}{RDU}& \multirow{2}*{Params}  & \multirow{2}*{MAdds} & \multirow{2}*{PSNR/SSIM}\\
    \cline{2-3}
    &Conv-type  & Rep-method & &  \\
    \midrule
    \multirow{8}*{\rotatebox{90}{DDistill-SR}}
    &Static &Static& 493K & 26.76G     & 28.21/0.8129\\ 
    \cline{2-6}
    &\multirow{3}*{DYConv}&Static& 1596K & 27.65G   &28.22/0.8132    \\
    & &RepVGG& 1758K  & 30.01G   &28.22/0.8133     \\
    & &DBB & -  & -   & -   \\
    \cline{2-6}
    &\multirow{3}*{DCD}&Static & 675K  & 32.62G    & 28.28/0.8140\\ 
    &&RepVGG & 720K   & 34.97G   & 28.29/0.8140\\ 
    &&DBB   & 1173K   & 60.80G  & 28.32/0.8148\\ 
    \midrule
    \multirow{5}*{\rotatebox{90}{RFDN}} 
    &Static &Static& 582K& 31.72G  & 28.24/0.8130    \\
    \cline{2-6}
    &\multirow{2}*{DYConv}&Static  & 1704K& 31.72G  &28.26/0.8133    \\
    &&RepVGG  & 1870K& 35.21G  & {28.26}/0.8137    \\
    \cline{2-6}
    &\multirow{2}*{DCD}&Static& 716K& 33.17G   & 28.31/0.8147 \\
    &&DBB      & 1224K& 61.98G   & {28.33}/{0.8153}    \\ 
    \bottomrule
  \end{tabular}
\label{tab:ablation_RDU}
\end{table}

\begin{figure}[!t]
  \vspace{0.1 cm}
  \setlength\tabcolsep{1.2pt} 
  \centering
  \scriptsize
  \begin{tabular}{ccccc}
    \includegraphics[width=.19\linewidth, height=1.8cm]{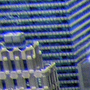} & \includegraphics[width=.19\linewidth, height=1.8cm]{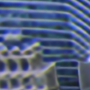} & \includegraphics[width=.19\linewidth, height=1.8cm]{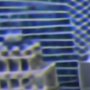}& \includegraphics[width=.19\linewidth, height=1.8cm]{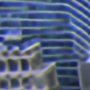}& \includegraphics[width=.19\linewidth, height=1.8cm]{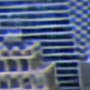} \\
   \includegraphics[width=.19\linewidth, height=1.8cm]{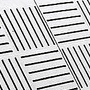} & \includegraphics[width=.19\linewidth, height=1.8cm]{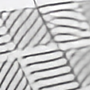} & \includegraphics[width=.19\linewidth, height=1.8cm]{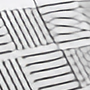}& \includegraphics[width=.19\linewidth, height=1.8cm]{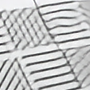}& \includegraphics[width=.19\linewidth, height=1.8cm]{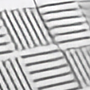} \\
   HR & RFDN & RFDN w/ RDU& Ours w/o DDF& Ours\\

  \end{tabular}
  \caption{Visual results on Urban100 ($\times$4)~\cite{Urban100}. The models with the proposed RDU and DDF better handle the structural distortion.}
  \label{tab:AblationStudy_fig}
  \end{figure}

\subsection{Ablation Studies}
In this section, a series of ablation studies are conducted to examine \B{the effectiveness of each component of} DDistill-SR, including the RDU module and the DDF module. 
We totally train the involved methods over 500k iterations for fairness. The multi-adds and running time are counted when the output size is $1280\times720$ for all scales.

\subsubsection{Effectiveness of the RDU}
We examine the effectiveness of the proposed RDU by investigating its static and dynamic representation capability, respectively.

\textbf{Reparameterization}.
The invariance of the reparameterizing procedure is first studied to \B{ensure} that the technique is harmless for running-time performance. We use DBB-style RepConv onto the DDistill-SR to measure the \B{effects} of reparameterization.
As shown in Table~\ref{tab:ablation_repara}, the evaluated difference between training and inference modules is less than 0.1\textperthousand, which implies the equivalence of performances. 
Furthermore, the inference model gains a slight lead (0.05\,dB on PSNR) while maintaining the parameter number and running time as the w/o Rep model.

\begin{table}[!t]
  \centering
  \caption{Investigations of \B{the} DDF module on Manga109~\cite{manga109} with $\times 4$ \textbf{BI} degradation.}
  \small
  \setlength\tabcolsep{3.0pt}
  \begin{tabular}{lcccc}
    \toprule
    Model & Params  & MAdds & FLOPs & PSNR/SSIM \\
    \midrule
    Baseline & 493K & 26.76G & 26.97G   & 30.49/0.9085\\ 
    RDU (w/o Rep) & 596K &28.18G & 28.33G& 30.49/0.9083    \\
    RDU (w/o Rep) + DDF  & 675K  & 32.62G &  32.88G & 30.61/0.9095\\ 
    RDU + DDF  & 675K  & 32.62G &  32.88G & 30.79/0.9098\\ 
    \bottomrule
  \end{tabular}
\label{tab:ablation_DDF}
\end{table}

\begin{figure}[!t]
  \flushleft
  \includegraphics[width=3.2in]{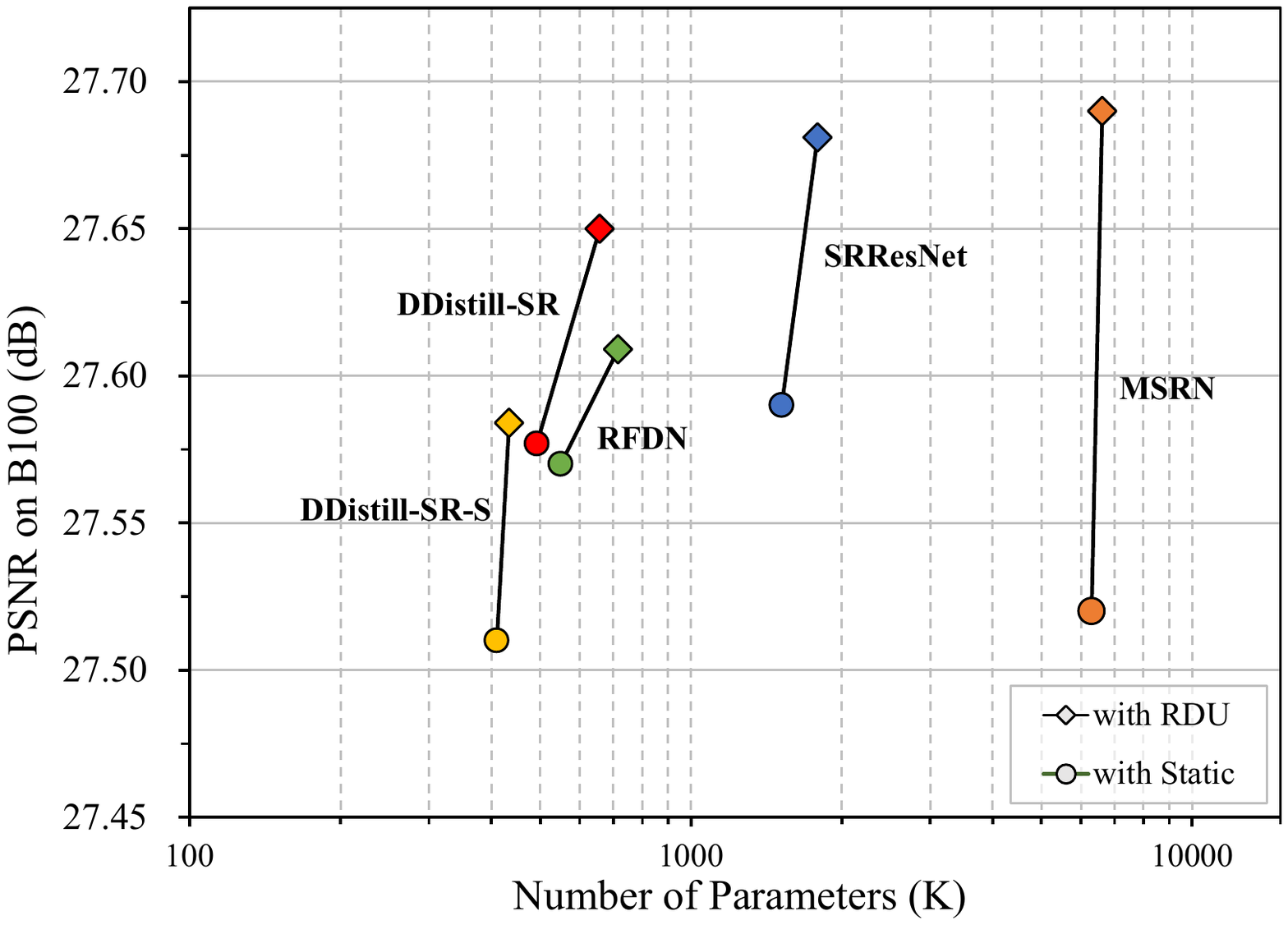} 
  \caption{Illustration of the PSNR vs. model parameters between RDU and static convolution with different models on B100~\cite{B100} for the scale factor $\times 4$. The diamond marks \B{denote} models with RDUs and the round marks denote \B{the} static convolution.}
  \label{fig_net_unit}
 \end{figure}

\textbf{Dynamic convolution}.
We further investigate the performance of two dynamic convolution methods, the traditional DYConv based on Eq.~\ref{dyconv} and the advanced DCD based on Eq.~\ref{dcd}. Specifically, we test the w/o Rep models in this part to show the separate potentiality of dynamic convolution on SR tasks.
Compared with static convolution, both DYConv and DCD can limitedly boost super-resolution tasks but bring larger training sizes, where \B{the number of} parameters \B{increases} to 1596k/596k and \B{the} computations increase to 1.42G/1.45G.

\textbf{Integrated RDU}.
\B{To answer the question of} ``How to best integrate these two techniques?", we implement RDUs by inserting different RepConvs (\eg, RepVGG and DBB) into dynamic convolution (\eg, DYConv and DCD) and investigate their validity in Table~\ref{tab:ablation_RDU} and Fig.~\ref{tab:AblationStudy_fig}. For intuition and convenience, we use Conv-Type/Rep-Method to enumerate possible implementations of RDU. 
Generally, models with integrated RDU outperform the static baseline while sharing similar deployment \B{costs}.
For the same reparameterization method, RDU using DCD leads 0.07\,dB-0.1\,dB ahead of traditional DYConv on both DDistill-SR and RFDN. In addition, the complicate DBB leads RepVGG by 0.04\,dB. Hence, the RDU with DCD/DBB can obtain better improvements. Besides, we note that the complex reparameter block also increases the training burden (\eg, the DYConv/DBB is unable to train).
To balance the training cost and the final restored quality, we adopt \B{the} DCD/DBB type RDU in our DDistill-SR.

\textbf{RDU on other models}.
We validate the effectiveness of RDU by leveraging it in other SR methods, including SRResNet and MSRN. 
The results in Fig.~\ref{fig_net_unit} show that the RDU improves PSNR over all these models with \B{few} extra parameters, especially for networks with deep feature extraction stage. 
In SRResNet and MSRN, the PSNR scores increase by 0.1\,dB and 0.17\,dB \B{when we simply replace} the vanilla convolution layer with our RDU.
Nevertheless, the improvement through individually using RDU in DDistill-SR is not as conspicuous as expected. This might be caused by the shallow structure since only 4 backbone blocks are stacked in our network for lightness, while 6 blocks of similar sizes are utilized in RFDN and IMDN.

\subsubsection{DDF Module}
To solve \B{this} problem, we introduce the DDF module with less additional cost to extend the dynamic representations in the local and global ranges.
As shown in Table~\ref{tab:ablation_DDF}, DDistill-SR with DDF achieves remarkably higher quantitative \B{indices}, \B{which leads} to \B{approximately} 0.3\,dB improvement over the baseline. Meanwhile, we discuss the critical role of DDF in models equipping RDU (w/o Rep), where DDF raises the PSNR index by 0.12\,dB. 
We also present the visual expression in Fig.~\ref{tab:AblationStudy_fig}, \B{where} the DDF helps restore the direction of lines in building pictures. It can be empirically found that our DDF module improves the restoration quality by a large margin.

\begin{table}[!t]
 \centering
 \caption{Investigations the effect of latent space $L$ in RDU. DDistill-SR with 56 channels and Dynamic Fusion module is adopted on 5 \textbf{BI} degraded testsets with scale factor $\times 4$.}
 \small
 \setlength\tabcolsep{15pt}
 \begin{tabular}{lrcc}
  \toprule
   $L$      &      Params&      MAdds&      PSNR/SSIM\\
   \midrule

   0 & 493K & 26.76G &28.21/0.8129\\
   8 & 588K & 31.87G &28.28/0.8143\\
   16 & 675K & 32.62G &28.32/0.8148\\
   24 & 830K & 33.29G &28.35/0.8153\\
   \bottomrule
  \end{tabular}
\label{tab:ablation_Latent_Space}
\end{table}

\subsubsection{Effectiveness of Different Latent Spaces}
We now balance the performance with the latent space $L$ of DCD branches.
Table~\ref{tab:ablation_Latent_Space} shows the impact of different values of $L$ on the model size and evaluated index. 
$L$ is a hyper-parameter to determine the latent space of DCD. 
The results imply that both performance and size increase with the growth of $L$, where the performance \B{increases at a} much slower \B{speed} than \B{the} model size.
To achieve an intuitive trade-off relationship, we finally set $L = 16$ in DDistill-SR and $L = 8$ in DDistill-SR-S.

\subsubsection{Effectiveness of Different RDU Architectures}
\begin{figure}[!t]
  \centering
  \includegraphics[width=3.3in]{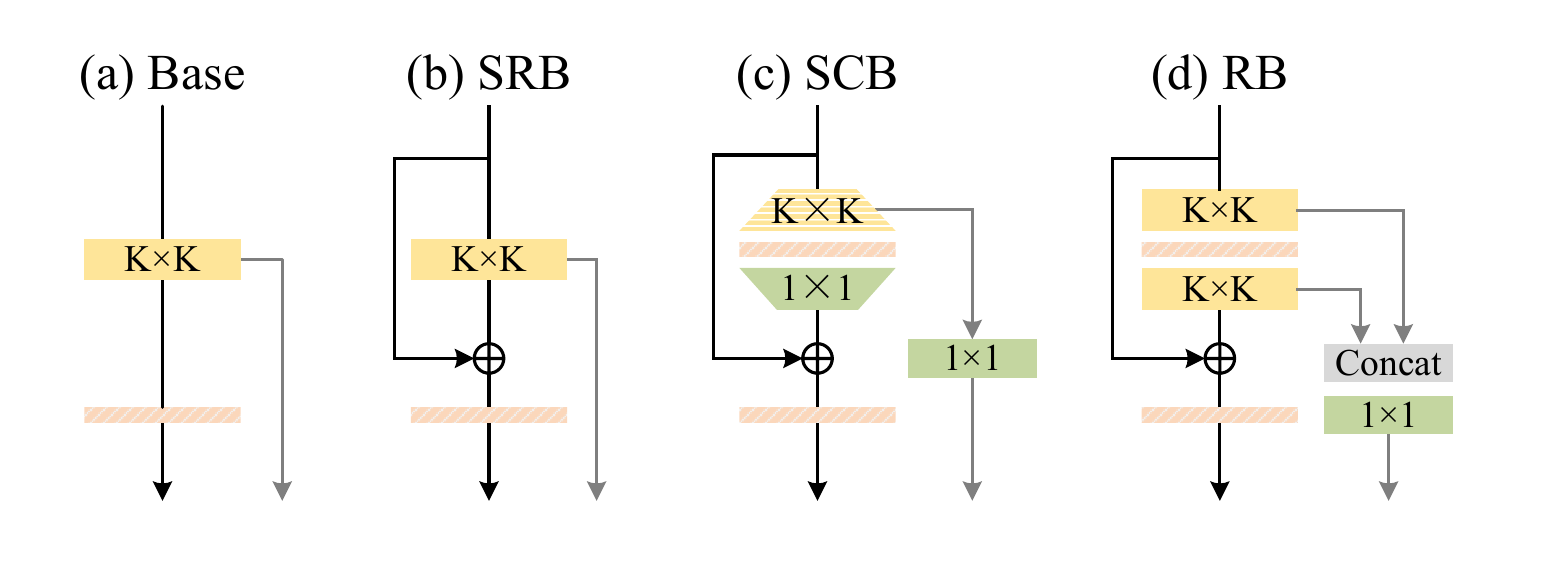} 
  \caption{\B{Illustration of} the four potential ways to implement RDU with diverse architectures: (a) base block (Base), (b) shallow residual block (SRB), (c) separable convolution block (SCB) and (d) residual block (RB). The basic $K\times K$ block represents Dynamic RepConv. We adopt (a) in DDistill-SR implementation and (c) in \B{DDistill-SR-S}.}
  \label{fig_RDU2}
 \end{figure}

We implement RDU with four common connection modes as shown in Fig.~\ref{fig_RDU2}.
The quality of restored images is evaluated on \textbf{BI} testsets ($\times 2$) in Table~\ref{tab:ablation_arch}.  
Specifically, DDistill-SR with SCB achieves comparative performance with 31\% multi-adds and nearly the same model size as the reference method. 
The Base and SRB structure gain an increase of about 0.16\,dB with diminished 58K parameters and 32.3G computations compared to IMDN on $\times 2$ testsets. 
RB further improves the restoration qualities with higher cost on \B{the} model scale.
These compositions can adapt to different tasks, and replace standard static blocks, \B{such as} residual and separable convolution. 
\B{During} our experiments, we apply the Base architecture to DDistill-SR \B{due to} its outperforming and convenience and adopt SCB in DDistill-SR-S for its lightness. 

\begin{table}[!t]
  \centering
  \small
  \setlength\tabcolsep{12pt}
  \caption{Evaluation of different architectures in RDU. Experiments are performed on five \textbf{BI} sets ($\times 2$) by average.}
  \begin{tabular}{lrrc}
    \toprule
    Method       & Params   & MAdds    & PSNR/SSIM\\%
    \midrule
    IMDN~\cite{IMDN}      & 715K     & 158.7G   & 34.55/0.9359\\ 
    Base       & 657K     & 126.4G   & 34.69/0.9368\\
    SRB        & 657K     & 126.4G   & 34.70/0.9368\\
    SCB        & 580K     & 38.1G    & 34.57/0.9363\\
    RB         & 1129K    & 216.2G   & 34.74/0.9373 \\
    \bottomrule
  \end{tabular}
  \label{tab:ablation_arch}
\end{table}

\begin{table*}[!t]
  \centering
  \caption{Average PSNR/SSIM for scale $\times 2$, $\times 3$ and $\times 4$ on datasets Set5~\cite{set5}, Set14~\cite{set14}, B100~\cite{B100}, Urban100~\cite{Urban100}, and Manga109~\cite{manga109} with \textbf{BI} degradation. The best/second-best results are highlighted in \textbf{bold} and \underline{underlined}, respectively.}
  \small
  \setlength\tabcolsep{9pt}
  \begin{tabular}{lcrccccc}
  \toprule
  \multirow{2}{*}{Method} & \multicolumn{1}{l}{\multirow{2}{*}{Scale}} & \multirow{2}{*}{Params}  & Set5 & Set14 & B100 & Urban100 & Manga109 \\
  & \multicolumn{1}{l}{} &  &PSNR/SSIM & PSNR/SSIM & PSNR/SSIM & PSNR/SSIM & PSNR/SSIM \\
  \midrule
  SRCNN~\cite{SRCNN} &\multirow{13}{*}{$\times$2}& 8K & 36.66/0.9542 & 32.45/0.9067 & 31.36/0.8879 & 29.50/0.8946 & 35.60/0.9663 \\
  FSRCNN~\cite{FSRCNN} &  & 13K & 37.00/0.9558 & 32.63/0.9088 & 31.53/0.8920 & 29.88/0.9020 & 36.67/0.9710 \\
  VDSR~\cite{VDSR} &  & 666K & 37.53/0.9587 & 33.03/0.9124 & 31.90/0.8960 & 30.76/0.9140 & 37.22/0.9750 \\
  LapSRN~\cite{LapSRN} &  & 251K & 37.52/0.9591 & 32.99/0.9124 & 31.80/0.8952 & 30.41/0.9103 & 37.27/0.9740 \\
  MemNet~\cite{MemNet} &  & 678K & 37.78/0.9597 & 33.28/0.9142 & 32.08/0.8978 & 31.31/0.9195 & 37.72/0.9740 \\
  IDN~\cite{IDN} &  & 553K & 37.83/0.9600 & 33.30/0.9148 & 32.08/0.8985 & 31.27/0.9196 & 38.01/0.9749 \\
  CARN~\cite{CARN} &  & 1592K & 37.76/0.9590 & 33.52/0.9166 & 32.09/0.8978 & 31.92/0.9256 & 38.36/0.9765 \\
  LESRCNN~\cite{LESRCNN} &  & - &  37.65/0.9586 & 33.32/0.9148 & 31.95/0.8964  & 31.45/0.9206 & 38.09/0.9759 \\
  IMDN~\cite{IMDN} &  & 694K & {38.00}/{0.9605} & {{33.63}}/0.9177 & \underline{32.19}/0.8996 & {32.17}/{0.9283} & {38.88}/{0.9774} \\
  PAN~\cite{PAN}& & 261K & {38.00/0.9605} & {{33.59/0.9181}} & {32.18/{0.8997}} & {{32.01}/{0.9273}} & {{38.70}/0.9773} \\  
  RFDN~\cite{RFDN}& & 534K & \underline{38.05/0.9606} & \underline{33.68/0.9184} & 32.16/{0.8994} & {32.12}/{0.9278} & {38.88}/0.9773 \\
  \textbf{DDistill-SR-S (Ours)}& & 414K & {38.03}/\underline{0.9606}&{33.61}/{0.9182} & \underline{32.19/0.9000}& \underline{32.18/0.9286} & \underline{38.94/0.9777} \\
  \textbf{DDistill-SR (Ours)}& & 657K & \textbf{38.08/0.9608} & \textbf{33.73/0.9195} & \textbf{32.25/0.9007} & \textbf{32.39/0.9301} & \textbf{39.16/0.9781} \\
  \midrule
  SRCNN~\cite{SRCNN} &\multirow{13}{*}{$\times$3}&  8K & 32.75/0.9090 & 29.30/0.8215 & 28.41/0.7863 & 26.24/0.7989 & 30.48/0.9117 \\
  FSRCNN~\cite{FSRCNN} &  & 13K &33.18/0.9140 & 29.37/0.8240 & 28.53/0.7910 & 26.43/0.8080 & 31.10/0.9210 \\
  VDSR~\cite{VDSR} &  & 666K & 33.66/0.9213 & 29.77/0.8314 & 28.82/0.7976 & 27.14/0.8279 & 32.01/0.9340 \\
  LapSRN~\cite{LapSRN} &  & 502K & 33.81/0.9220 & 29.79/0.8325 & 28.82/0.7980 & 27.07/0.8275 & 32.21/0.9350 \\
  MemNet~\cite{MemNet} &  & 678K & 34.09/0.9248 & 30.00/0.8350 & 28.96/0.8001 & 27.56/0.8376 & 32.51/0.9369 \\
  IDN~\cite{IDN} &  & 553K & 34.11/0.9253 & 29.99/0.8354 & 28.95/0.8013 & 27.42/0.8359 & 32.71/0.9381 \\
  CARN~\cite{CARN} &  & 1592K &  34.29/0.9255 & 30.29/0.8407 & 29.06/0.8034 & 28.06/0.8493 & 33.50/0.9440 \\
  {LESRCNN~\cite{LESRCNN}} &  & {-} &  {33.93/0.9231} &{30.12/0.8380} &{28.91/0.8005} & {27.70/0.8415} &{ 32.91/0.9402} \\
  IMDN~\cite{IMDN} &  & 703K & 34.36/0.9270 & 30.32/{0.8417} & {29.09}/0.8046 & 28.17/0.8519 & 33.61/0.9445 \\
  {PAN~\cite{PAN}} &  & {261K} & {34.40/0.9271} & {\underline{30.36/0.8423}} & {\underline{29.11}/0.8050} & {28.11/0.8511} & {33.61/0.9448} \\
  RFDN~\cite{RFDN}& & 541K & \underline{34.41}/{0.9273} & {30.34/0.8420} & {29.09}/{0.8050} & \underline{28.21}/{0.8525} & {33.67}/{0.9449} \\
  \textbf{DDistill-SR-S (Ours)}& & 414K & {34.37}/\underline{0.9275} & {30.34/0.8420}& \underline{29.11/0.8053} & {28.19}/\underline{0.8528} & \underline{33.69/0.9451}\\
  \textbf{DDistill-SR (Ours)}& & 665K & \textbf{34.43/0.9276} & \textbf{30.39/0.8432} & \textbf{29.16/0.8070} & \textbf{28.31/0.8546} & \textbf{33.97/0.9465} \\ 
  \midrule
  SRCNN~\cite{SRCNN} & \multirow{13}{*}{$\times$4} & 8K & 30.48/0.8626 & 27.50/0.7513 & 26.90/0.7101 & 24.52/0.7221 & 27.58/0.8555 \\
  FSRCNN~\cite{FSRCNN} &  & 13K & 30.72/0.8660 & 27.61/0.7550 & 26.98/0.7150 & 24.62/0.7280 & 27.90/0.8610 \\
  VDSR~\cite{VDSR} &  & 666K & 31.35/0.8838 & 28.01/0.7674 & 27.29/0.7251 & 25.18/0.7524 & 28.83/0.8870 \\
  LapSRN~\cite{LapSRN} &  & 502K &  31.54/0.8852 & 28.09/0.7700 & 27.32/0.7275 & 25.21/0.7562 & 29.09/0.8900 \\
  MemNet~\cite{MemNet} &  & 678K & 31.74/0.8893 & 28.26/0.7723 & 27.40/0.7281 & 25.50/0.7630 & 29.42/0.8942 \\
  IDN~\cite{IDN} &  & 553K &  31.82/0.8903 & 28.25/0.7730 & 27.41/0.7297 & 25.41/0.7632 & 29.41/0.8942 \\
  CARN~\cite{CARN} &  & 1592K &  32.13/0.8937 & {28.60}/0.7806 & {27.58}/0.7349 & 26.07/0.7837 & 30.47/{0.9084} \\
  {LESRCNN~\cite{LESRCNN} }&  & {-} &  {31.88/0.8903} & {28.44/0.7772} & {27.45/0.7313} & {25.77/0.7732} & {30.01/0.9017} \\
  IMDN~\cite{IMDN} &  & 715K & {32.21}/{0.8948} & 28.58/0.7811 & 27.56/0.7353 & 26.04/0.7838 & 30.45/0.9075 \\
  {PAN~\cite{PAN}} &  & {272K} & {32.13/0.8948} & {28.61/0.7822} & {\underline{27.59}/0.7363} & {26.11/0.7854} & {30.51/\underline{0.9095}}\\
  RFDN~\cite{RFDN}& & 550K & \underline{32.24}/{0.8952} & {28.61}/{0.7819} & {27.57}/{0.7360} & {26.11}/{0.7858} & \underline{30.58}/{0.9089} \\
  \textbf{DDistill-SR-S (Ours)}& & 434K& {32.23}/\underline{0.8960} & \underline{28.62/0.7823} & 27.58/\underline{0.7365} & \underline{26.20/0.7891} & {30.48}/{0.9090}\\
  \textbf{DDistill-SR (Ours)}& & 675K& \textbf{32.29/0.8961} &\textbf{28.69/0.7833}& \textbf{27.65/0.7385}&\textbf{26.25/0.7893}&\textbf{30.79/0.9098}\\
  \bottomrule
  \end{tabular}
  \label{tab:sota_BI}
\end{table*}

\subsection{Comparison with State-of-the-Art Models}

\subsubsection{Results with \textbf{BI} Degradation Method}
To evaluate the learning ability of the proposed method, we compare DDistill-SR and DDistill-SR-S with several lightweight SR models as shown in Table~\ref{tab:sota_BI}, including SRCNN~\cite{SRCNN}, FSRCNN~\cite{FSRCNN}, VDSR~\cite{VDSR}, LapSRN~\cite{LapSRN}, MemNet~\cite{MemNet}, IDN~\cite{IDN}, CARN~\cite{CARN}, {LESRCNN~\cite{LESRCNN}}, IMDN~\cite{IMDN}, {PAN~\cite{PAN}}, and RFDN~\cite{RFDN}. 
We test these methods on the Set5, Set14, B100, Urban100, and Manga109 datasets with magnification factors of $\times2$, $\times3$, and $\times4$.
DDistill-SR clearly demonstrates a noticeable improvement in PSNR and SSIM compared to all competing methods. 
Here, DDistill-SR-S achieves comparable or better results with only 434k parameters and 24.2G MAdds computation, and DDistill-SR outperforms all state-of-the-art methods. In Manga109, the average PSNR improvement of DDistill-SR over RFDN is 0.28\,dB, 0.30\,dB, and 0.21\,dB for three upscaling factors, respectively. Moreover, experiments verify the effectiveness and efficiency of DDistill-SR in terms of the parameter size and computation cost. In detail, for the $\times4$ SR task, DDistill-SR-S uses 116k and 181k fewer parameters than IMDN and RFDN, respectively. Compared to the recently proposed PAN with only 272k parameters and 28.2G MAdds, our Distill-SR-S has advantages in calculation and restoration quality. For the $\times 2$ task, our method advances PAN on all datasets with a maximal 0.24\,dB leading. A similar result happens to LESRCNN, which requires fewer calculations but fails to maintain advantages in restoration quality.
Our DDistill-SR contains 675k parameters and 32.62G multi-adds operations, which achieves 0.11\,dB, 0.21\,dB and 0.34\,dB performance improvement over IMDN on Set14, Urban100 and Manga109 ($\times4$). Generally, our DDistill-SR models obtain better trade-offs between performance and applicability.

\begin{figure*}[!t]
  \setlength\tabcolsep{1.5pt}
  \centering
  \footnotesize
  \begin{tabular}{cccccc}
 \multirow{-6.60}{*}{\includegraphics[width=.24\linewidth, height=0.245\linewidth]{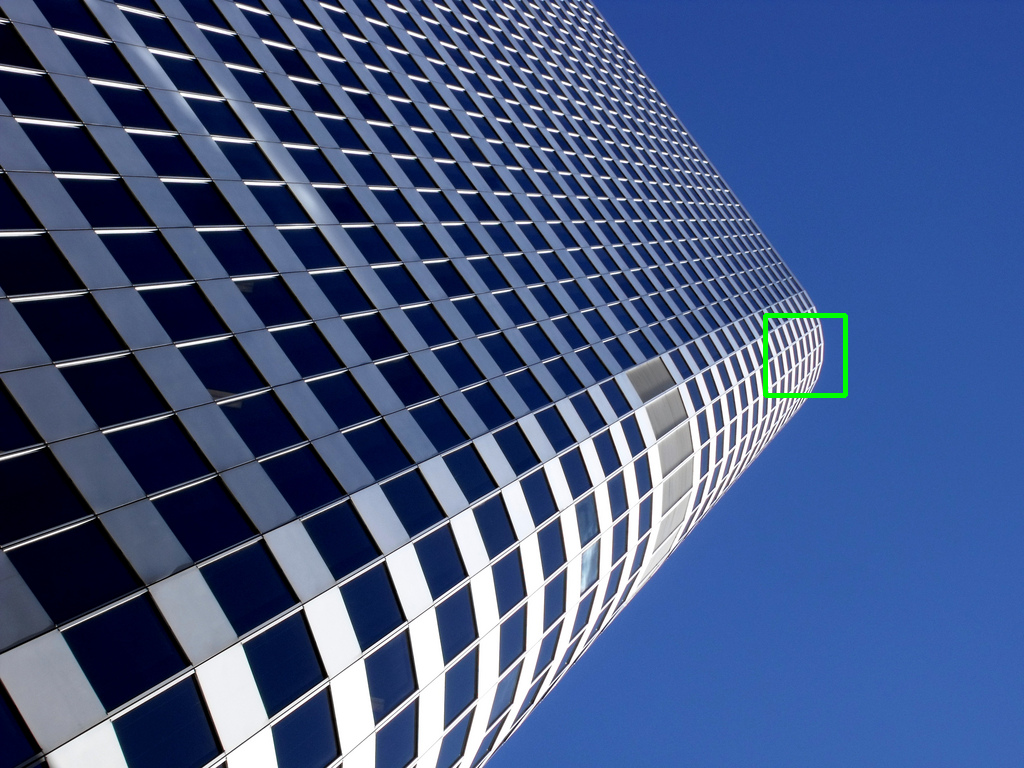}} 
   & \includegraphics[width=.13\linewidth, height=0.11\linewidth]{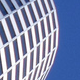} & \includegraphics[width=.13\linewidth, height=0.11\linewidth]{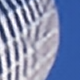} & \includegraphics[width=.13\linewidth, height=0.11\linewidth]{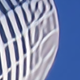}& \includegraphics[width=.13\linewidth, height=0.11\linewidth]{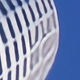}& \includegraphics[width=.13\linewidth, height=0.11\linewidth]{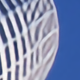} \\
   & HR PSNR & FSRCNN 25.82\,dB& VDSR 27.35\,dB & CARN 27.71\,dB&  {LESRCNN 27.50dB}  \\
   & \includegraphics[width=.13\linewidth, height=.11\linewidth]{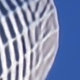} & \includegraphics[width=.13\linewidth, height=0.11\linewidth]{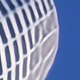} & \includegraphics[width=.13\linewidth, height=0.11\linewidth]{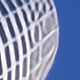} & \includegraphics[width=.13\linewidth, height=0.11\linewidth]{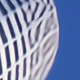}& \includegraphics[width=.13\linewidth, height=0.11\linewidth]{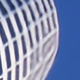} \\
   \emph{img\_012} from Urban100 & IMDN 27.37\,dB& {PAN 27.85\,dB} & RFDN \underline{28.03}\,dB & \textbf{Ours-S} 27.82\,dB &\textbf{Ours} \textbf{28.39}\,dB\\  
  \multirow{-6.6}{*}{\includegraphics[width=.24\linewidth, height=0.245\linewidth]{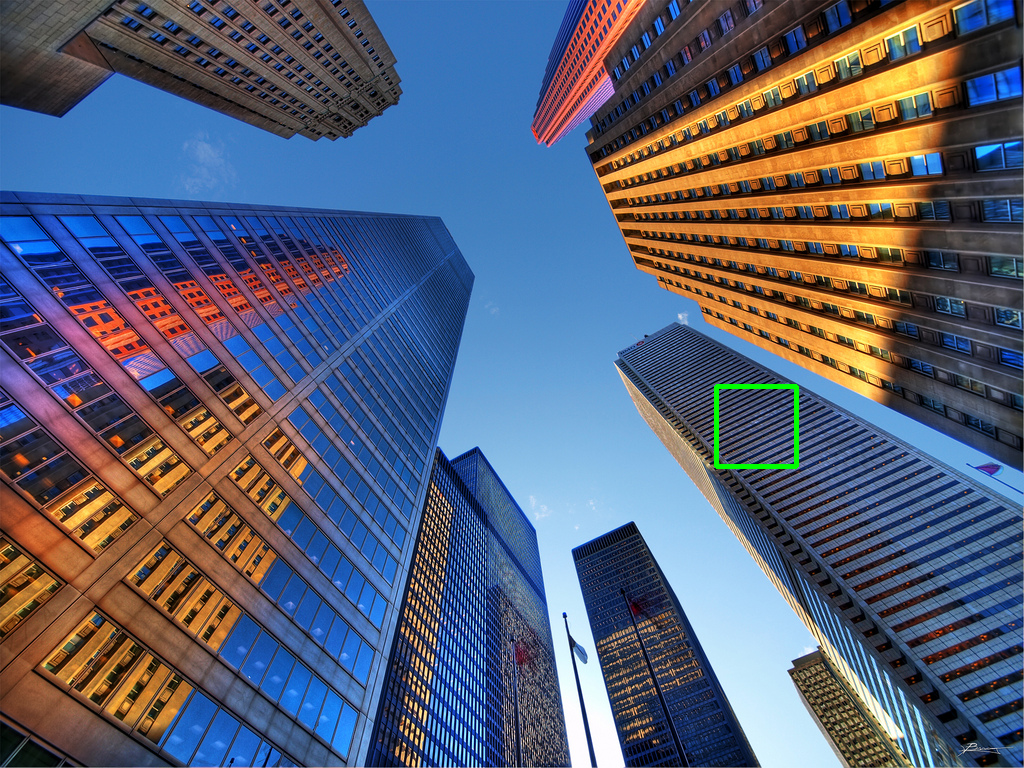}} 
   & \includegraphics[width=.13\linewidth, height=0.11\linewidth]{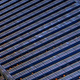} & \includegraphics[width=.13\linewidth, height=0.11\linewidth]{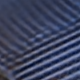} & \includegraphics[width=.13\linewidth, height=0.11\linewidth]{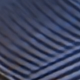}& \includegraphics[width=.13\linewidth, height=0.11\linewidth]{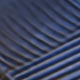}& \includegraphics[width=.13\linewidth, height=0.11\linewidth]{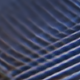} \\
   & HR PSNR & FSRCNN 23.25\,dB& VDSR 23.53\,dB & CARN 23.85\,dB& {LESRCNN 23.68dB}  \\
   & \includegraphics[width=.13\linewidth, height=0.11\linewidth]{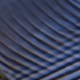} & \includegraphics[width=.13\linewidth, height=0.11\linewidth]{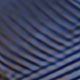} & \includegraphics[width=.13\linewidth, height=0.11\linewidth]{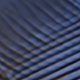} & \includegraphics[width=.13\linewidth, height=0.11\linewidth]{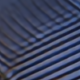}& \includegraphics[width=.13\linewidth, height=0.11\linewidth]{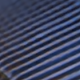} \\
   \emph{img\_012} from Urban100 & {IMDN 23.81\,dB}& {PAN 23.85\,dB} & {RFDN 23.88\,dB} & {\textbf{Ours-S} \underline{23.91}\,dB }&{\textbf{Ours} \textbf{23.97}\,dB }\\  
     \multirow{-6.6}{*}{\includegraphics[width=.24\linewidth, height=0.245\linewidth]{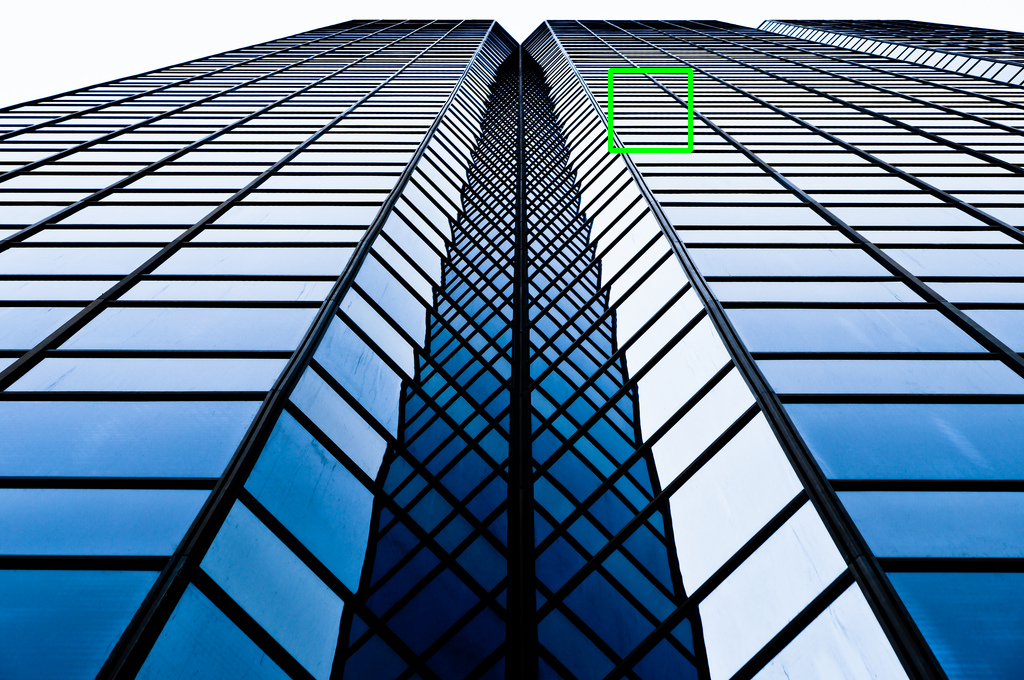}} 
   & \includegraphics[width=.13\linewidth, height=0.11\linewidth]{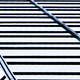} & \includegraphics[width=.13\linewidth, height=0.11\linewidth]{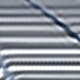} & \includegraphics[width=.13\linewidth, height=0.11\linewidth]{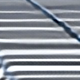}& \includegraphics[width=.13\linewidth, height=0.11\linewidth]{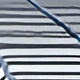}& \includegraphics[width=.13\linewidth, height=0.11\linewidth]{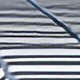} \\
   & {HR PSNR} & {FSRCNN 18.00\,dB }&{ VDSR 18.84\,dB }&{ CARN 19.42\,dB}& {LESRCNN 19.39dB}  \\
   & \includegraphics[width=.13\linewidth, height=0.11\linewidth]{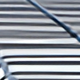} & \includegraphics[width=.13\linewidth, height=0.11\linewidth]{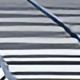} & \includegraphics[width=.13\linewidth, height=0.11\linewidth]{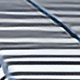} & \includegraphics[width=.13\linewidth, height=0.11\linewidth]{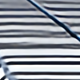}& \includegraphics[width=.13\linewidth, height=0.11\linewidth]{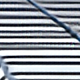} \\
   {\emph{img\_067} from Urban100} & {IMDN 19.64\,dB}& {PAN 19.59\,dB} & {RFDN 19.99\,dB} & {\textbf{Ours-S} \underline{20.05}\,dB }&{\textbf{Ours} \textbf{20.27}\,dB}\\   
     \multirow{-6.6}{*}{\includegraphics[width=.24\linewidth, height=0.245\linewidth]{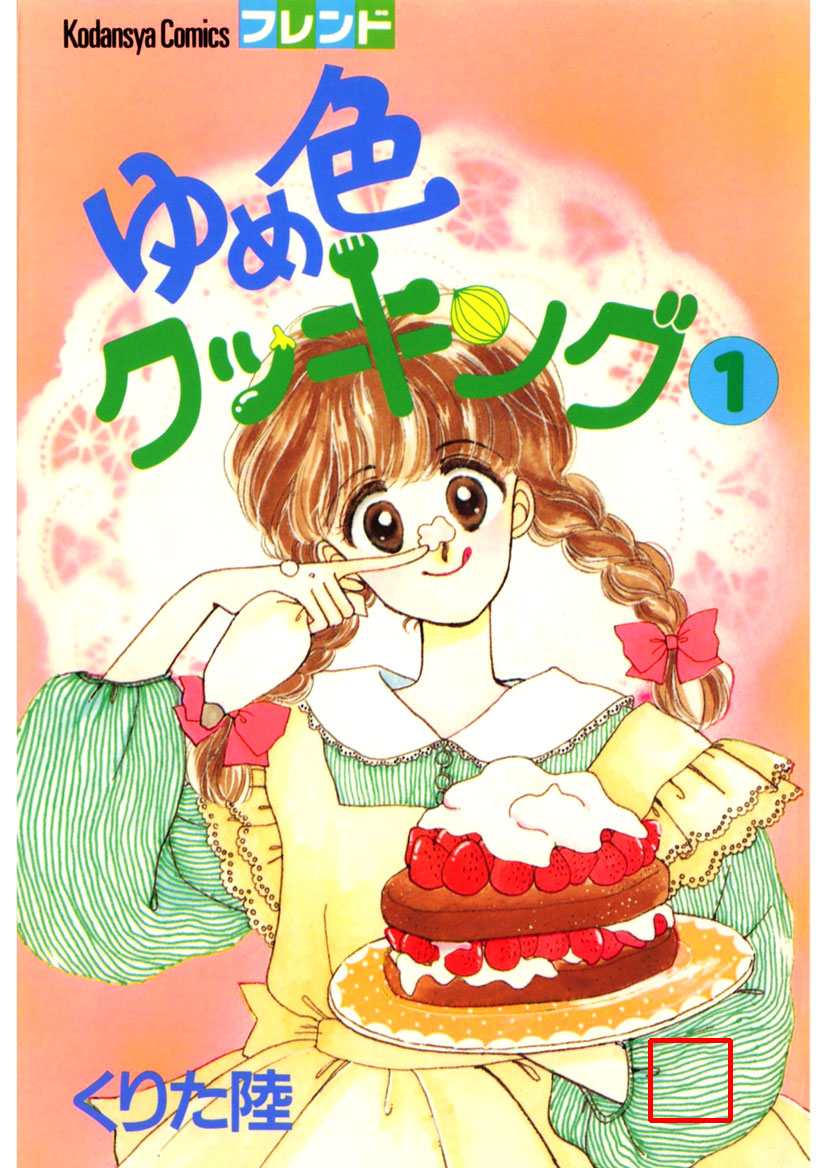}} 
   & \includegraphics[width=.13\linewidth, height=0.11\linewidth]{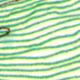} & \includegraphics[width=.13\linewidth, height=0.11\linewidth]{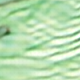} & \includegraphics[width=.13\linewidth, height=0.11\linewidth]{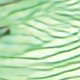}& \includegraphics[width=.13\linewidth, height=0.11\linewidth]{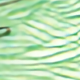}& \includegraphics[width=.13\linewidth, height=0.11\linewidth]{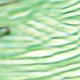} \\
   & HR-PSNR & FSRCNN 26.25\,dB& VDSR 27.39\,dB & CARN 27.62\,dB& {LESRCNN 27.72dB}  \\
   & \includegraphics[width=.13\linewidth, height=0.11\linewidth]{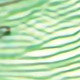} & \includegraphics[width=.13\linewidth, height=0.11\linewidth]{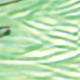} & \includegraphics[width=.13\linewidth, height=0.11\linewidth]{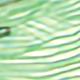} & \includegraphics[width=.13\linewidth, height=0.11\linewidth]{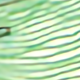}& \includegraphics[width=.13\linewidth, height=0.11\linewidth]{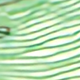} \\
   \emph{YumeiroCooking} from Manga109 & IMDN 27.89\,dB& {PAN 27.87\,dB} & RFDN 28.31\,dB & \textbf{Ours-S} \underline{28.32}\,dB &\textbf{Ours} \textbf{29.02}\,dB\\   
  \end{tabular}
  \caption{The comparison of DDistill-SR-S~(\textbf{Ours-S}) and DDistill-SR~(\textbf{Ours}) with other excellent algorithms. Our DDistill-SR family has better visual and qualitative results on the \textbf{BI} degradation dataset with scale factor $\times$4. The best/second-best PSNR results are \textbf{highlighted} and \underline{underlined}, respectively.}
  \label{tab:Bi_fig}
  \end{figure*}

In addition, we show the restored image and qualitative results in Fig.~\ref{tab:Bi_fig}. 
Our methods generally produce more accurate and vivid images than others. 
Specifically, DDistill-SR obtains sharp edges and precise contour, which reuses the dynamic residuals to maintain structural information. 
Meanwhile, DDistill-SR-S generates clear but disorganized figures, \B{since} DDistill-SR-S adopts RDU without DDF for lightweight delivery, which only strengthens \B{the} local information but ignores the network-level dynamic feature enhancement.
In \emph{img\_005}, the shape and orientation of building windows are not \B{correctly restored} in most existing methods except for our DDistill-SR. 
In terms of PSNR criteria, DDistill-SR beats the second-best method RFDN by 0.37\,dB.
The superiority is also corroborated in \emph{img\_012} and \emph{img\_067}, where DDistill-SR recovers structural direction from the interlaced and blurred LR images.
Referring to the last image \emph{YumeiroCooking}, DDistill-SR-S performs better than other methods. Compared to PAN, DDistill-SR-S obtains a 0.45\,dB improvement on PSNR, which is also reflected by the qualitative results where our DDistill-SR-S correctly \B{rebuilds} the stroke direction.
\B{In summary}, the HR images generated by DDistill-SR are closer to the ground truth than other state-of-the-art methods.

\subsubsection{Results with \textbf{BD} and \textbf{DN} Degradation Methods}
We further compare our DDistill-SR model with other SR methods~\cite{VDSR,SPMSR,IRCNN,SRMD} on the $\times3$ \textbf{BD} and \textbf{DN} degraded datasets. In Table~\ref{tab:BD_DN_sota}, we demonstrate the consistent performance improvement in PSNR and SSIM scores on the test sets Set5, Set14, B100, and Urban100. As a groundbreaking work, RDN~\cite{RDN} leverages dense connections between each layer to comprehensively utilize local layers under the \textbf{BD} and \textbf{DN} degradation. Our DDistill-SR method achieves similar performance with fewer parameters \B{than} RDN on \textbf{DN}. DDistill-SR is more suitable to extract meaningful features with certain noise \B{to introduce} a more effective representation module. Moreover, RDN contains 22M parameters, which is 33 times \B{more than the number of parameters of} DDistill-SR. For \textbf{BD} degradation, DDistill-SR achieves similar performance to high-cost RDN and far outperforms other low-cost networks. However, the gap between RDN and DDistill-SR is less than 0.09\,dB, which is much lower than the benefit of our approach over other methods. Compared with SRMDNF, our method improves PSNR and SSIM by 0.33\,dB, and 0.0061 on average for the four datasets.

We also show intuitive results to confirm the potential of our method in anti-warping and denoising. As depicted in Fig.~\ref{BI_DN}, for \emph{barbara} image with \textbf{BD} degradation, RDN and DDistill-SR remove blurring artifacts and recover the deficient structural edges.
However, our method better restores the pattern of the tablecloth and outperforms RDN in terms of sensory and evaluation \B{indices}. Moreover, DDistill-SR learns the missing high-frequency information from the highly aliased and noisy \textbf{DN} inputs and generates images with more details. 
These results clearly indicate that DDistill-SR is more effective than other lightweight methods. We attribute the lead to our enhancements on dynamic representation.

\begin{table*}[!t]
  \centering
  \small
  \setlength\tabcolsep{2.51pt}
  \caption{Average PSNR/SSIM for scale $\times 3$ on datasets Set5~\cite{set5}, Set14~\cite{set14}, B100~\cite{B100}, Urban100~\cite{Urban100} with \textbf{BD} and \textbf{DN} degradation. The best/second-best results are \textbf{highlighted} and \underline{underlined} respectively.}
  \begin{tabular}{cccr@{/}lcccccc}
    \toprule
             Dataset          & Type     &Bicubic     &\multicolumn{2}{c}{SPMSR~\cite{SPMSR}}   &VDSR~\cite{VDSR}&IRCNN\_G~\cite{IRCNN}&IRCNN\_C~\cite{IRCNN}&SRMDNF~\cite{SRMD}&RDN~\cite{RDN}&\textbf{DDistill-SR}\\%
    \midrule
    \multirow{2}{*}{Set5}     & \textbf{BD} &28.34/0.8161&32.21&0.9001                          &33.29/0.9139    &33.38/0.9182         &29.55/0.8246         &34.09/0.9242        &\textbf{34.57/0.9280}&\underline{34.39/0.9255}\\ 
                              & \textbf{DN} &24.14/0.5445&   -&-                                &27.42/0.7372    &24.85/0.7205         &26.18/0.7430         &27.74/0.8026        &\underline{28.46}/\textbf{0.8151}& \textbf{28.47}/\underline{0.8142}\\
    \midrule
    \multirow{2}{*}{Set14}    & \textbf{BD} &26.12/0.7106&28.89&0.8105                          &29.58/0.8259    &29.73/0.8292         &27.33/0.7135         &30.11/0.8364        &\textbf{30.53/0.8447}&\underline{30.44/0.8401}\\ 
                              & \textbf{DN} &23.14/0.4828&   -&-                                &25.60/0.6706    &23.84/0.6091         &24.68/0.6300         &26.13/0.6974        &\underline{26.60/0.7101}& \textbf{26.63/0.7120}\\
    \midrule
    \multirow{2}{*}{B100}     & \textbf{BD} &26.02/0.6733&28.13&0.7740                          &28.61/0.7900    &28.65/0.7922         &26.46/0.6572         &28.98/0.8009        &\textbf{29.23/0.8079}&\underline{29.19/0.8049}\\ 
                              & \textbf{DN} &22.94/0.4461&   -&-                                &25.22/0.6271    &23.89/0.5688         &24.52/0.5850         &25.64/0.6495        &\underline{25.93/0.6573}& \textbf{25.97/0.6607}\\
    \midrule
    \multirow{2}{*}{Urban100} & \textbf{BD} &23.20/0.6661&25.84&0.7856                          &26.68/0.8019    &26.77/0.8154         &24.89/0.7172         &27.50/0.8370        &\textbf{28.46/0.8581}&\underline{28.04/0.8463}\\ 
                              & \textbf{DN} &21.63/0.4701&   -&-                                &23.33/0.6579    &21.96/0.6018         &22.63/0.6205         &24.28/0.7092        &\underline{24.92/0.7362} & \textbf{24.93/0.7370} \\
    \bottomrule
  \end{tabular}
  \label{tab:BD_DN_sota}
\end{table*}

\begin{figure*}[!t]
  \setlength\tabcolsep{2.0pt} 
  \centering
  \scriptsize
  \begin{tabular}{cccccc}
    \multirow{-6.7}{*}{\includegraphics[width=.23\linewidth, height=4cm]{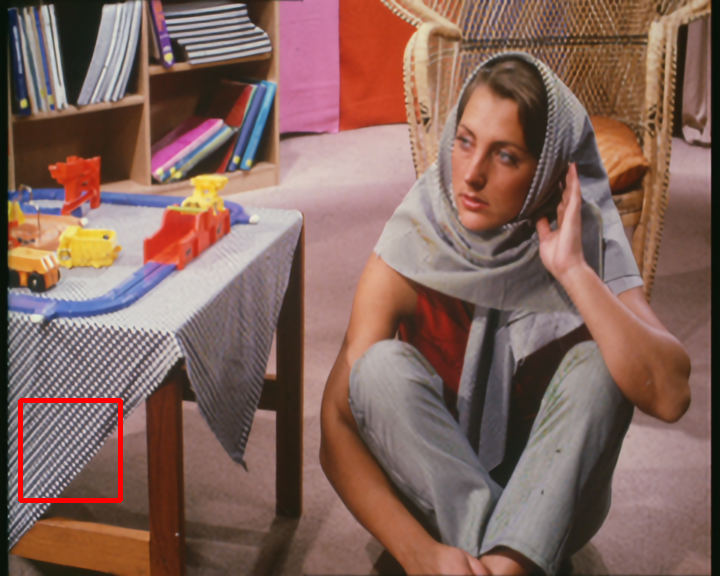}} 
    & \includegraphics[width=.12\linewidth, height=1.8cm]{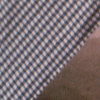} & \includegraphics[width=.12\linewidth, height=1.8cm]{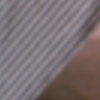}
    & \includegraphics[width=.12\linewidth, height=1.8cm]{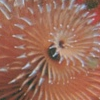} & \includegraphics[width=.12\linewidth, height=1.8cm]{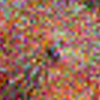}    &\multirow{-6.7}{*}{\includegraphics[width=.23\linewidth, height=4cm]{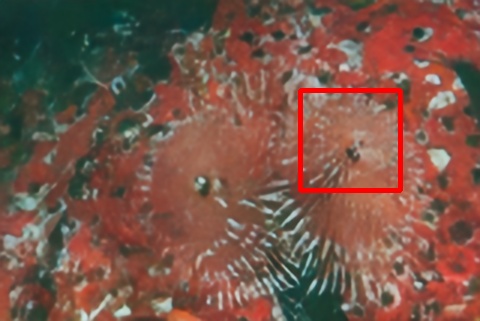}} \\
    & HR PSNR & Bicubic 25.24\,dB & HR PSNR & Bicubic 23.41\,dB &\\
    & \includegraphics[width=.12\linewidth, height=1.8cm]{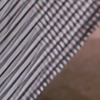} & \includegraphics[width=.12\linewidth, height=1.8cm]{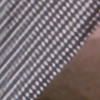} 
    &\includegraphics[width=.12\linewidth, height=1.8cm]{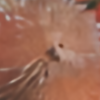} & \includegraphics[width=.12\linewidth, height=1.8cm]{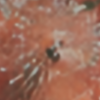} & \\
    \emph{barbara} from Set14~(\textbf{BD}) & RDN \underline{27.08}\,dB& \textbf{Ours} \textbf{27.24}\,dB & 
     RDN \underline{25.64}\,dB& \textbf{Ours} \textbf{25.80}\,dB &\emph{12084} from B100~(\textbf{DN}) \\
  \end{tabular}
  \caption{The comparison of DDistill-SR~(\textbf{Ours}) and RDN~\cite{RDN} on \textbf{BD} and \textbf{DN} degradation~($\times$3).}
  \label{BI_DN}
  \end{figure*}

\subsection{Pros and Cons}
In this paper, we propose a plug-in convolutional block RDU and \B{an} efficient dynamic distillation fusion module, which show better representations in lightweight frameworks. Although substantial improvement can be obtained in our work for lightweight SR, critical points \B{remain} to research prior to practical application. \B{First}, despite the proposed RDU \B{that} efficiently \B{combines} dynamic convolution and reparameterization strategy to capture more powerful feature extraction capabilities, extremely high overhead is unavoidable in the training process. It is necessary to reduce the training-time memory cost for deployment on large models. \B{Second}, the DDF module can collect and reuses the immediate features to erect the long-distance correlation for better quality. However, due to the extra parameters and calculations, the DDF is removed in the DDistill-SR-S. Recursive learning may \B{help} reduce the additional cost by sharing the DDF module among different levels. \B{By} overcoming \B{these} two points, we believe DDistill-SR can achieve better performance and applicability.

\section{Conclusion}\label{section:conclusion}
In this article, we propose an efficient lightweight super-resolution network, \B{called} DDistill-SR, to restore low-resolution images under limited time and space overhead. 
\B{To improve the} representation capacity in limited depth, we employ more powerful convolutional layers and more effectual fusion modules to fully use the static and dynamic signals in the distillation super-resolution framework. 
Hence, a novel reparameterized and dynamical RDU is proposed to enrich the extraction and distillation of high-frequency information. 
Extensive evaluations suggest that the proposed RDU may help with other lightweight vision applications.
Moreover, an efficient distillation block is devised to ensure lightweight and flexible feature processing. Furthermore, we conduct a dynamic fusion module to collect local and global dynamic residues to assist the final restoration. 
With these modifications, our DDistill-SR can adapt to different magnifications and degradation. Extensive evaluations well demonstrate that our DDistill-SR achieves the SOTA performance with relatively fewer parameters and calculations.


\ifCLASSOPTIONcaptionsoff
  \newpage
\fi

\bibliographystyle{IEEEtran}
\bibliography{IEEEexample}

\end{document}